\documentclass[preprint,11pt,superscriptaddress,amsmath,amssymb,nofootinbib,longbibliography]{revtex4-1}
	\usepackage{amssymb}
	\usepackage{amsmath}
	\usepackage{hyperref}
	\usepackage{graphicx}
	\usepackage{color}
	\newcommand{\nn}{\nonumber}
	\newcommand{\lam}{\lambda}
	\newcommand{\rbra}[1]{\left(#1\right)}
	\newcommand{\sbra}[1]{\left[#1\right]}
	\newcommand{\cbra}[1]{\left\{#1\right\}}
	\newcommand{\eval}[1]{\langle#1\rangle}
	\newcommand{\re}{\textrm{Re}}
	\newcommand{\im}{\textrm{Im}}
	\newcommand{\Tr}{\textrm{Tr}}
	\newcommand{\eq}[1]{\begin{align}#1\end{align}}
	\newcommand{\texteq}[1]{$#1$}
	\newcommand{\primed}[1]{{#1'}}

\allowdisplaybreaks[4]

\begin{document}

\title{Aligned CP-violating Higgs sector canceling the electric dipole moment}

%%%%%     authors     %%%%%
\author{Shinya Kanemura}
\email{kanemu@het.phys.sci.osaka-u.ac.jp}
\author{Mitsunori Kubota}
\email{mkubota@het.phys.sci.osaka-u.ac.jp}
\author{Kei Yagyu}
\email{yagyu@het.phys.sci.osaka-u.ac.jp}
\affiliation{Department of Physics, Osaka University, Toyonaka, Osaka 560-0043, Japan}

%%%%%     date     %%%%%
%\date{\today}

%%%%%     abstract     %%%%%
\begin{abstract}
We discuss the effect of CP violation in the aligned scenario of the general two-Higgs-doublet model, in which the Higgs potential and the Yukawa interaction provide additional CP-violating phases.
An alignment is imposed to the Yukawa interaction in order to avoid dangerous flavor changing neutral currents.
The Higgs potential is also aligned such that the coupling constants of the lightest Higgs boson, which is identified as the discovered Higgs boson with the mass of 125 GeV, are the same as those of the standard model.
In general, CP-violating phases originated by the Yukawa interaction and the Higgs potential are strongly constrained by the current data for the electric dipole moment (EDM).
It is found that in our scenario contributions from the two sources of CP violation can be destructive and consequently their total contribution can satisfy the EDM results, even when each CP-violating phase is large.
Such a large CP-violating phase can be tested at collider experiments by looking at the angular distributions of particles generated by the decays of the additional Higgs bosons.
\end{abstract}

%%%%%     preprint     %%%%%
\preprint{OU-HET-1048}

%%%%%%%%%%
%%%%%%%%%%
\maketitle

%%%%%     table of contents     %%%%%
%\tableofcontents
%\newpage

%%%%%%%%%%     introduction     %%%%%%%%%%
\section{Introduction}
In spite of the success of the standard model (SM) for elementary particles, there are still phenomena which cannot be explained by this model.
Baryon asymmetry of the Universe (BAU) is one of such phenomena beyond the SM.
It is well known that the Sakharov's three conditions have to be satisfied for viable baryogenesis which explains the observed BAU from a baryon symmetric universe\cite{Sakharov:1967dj};
(1) existence of baryon number changing interaction,
(2) C and CP violation,
and (3) departure from thermal equilibrium.
In the SM, these conditions can be satisfied qualitatively by the sphaleron transition, the Kobayashi-Maskawa (KM) phase, and the strongly first-order electroweak phase transition (EWPT), respectively\cite{Kuzmin:1985mm}.
However, the magnitude of CP violation by the KM phase is numerically not sufficient to realize the observed BAU\cite{Huet:1994jb}.
Furthermore, it turned out that the EWPT is crossover with the measured value of the mass of the Higgs boson\cite{Kajantie:1996mn}.
Therefore, a new physics model has to be introduced to explain the BAU.

So far, various new physics scenarios for baryogenesis have been proposed such as those based on grand unification theories\cite{Yoshimura:1978ex,Weinberg:1979bt}, leptogenesis\cite{Fukugita:1986hr}, electroweak baryogenesis (EWBG)\cite{Kuzmin:1985mm} and so on.
Among these scenarios, EWBG requires new physics at the TeV scale, so that it can be tested at high-energy collider experiments as well as flavor experiments and astrophysical observations.
Thus, it is interesting and timely to consider EWBG.
For the successful scenario of EWBG, the Higgs sector is extended in order to obtain additional CP-violating phases and the strongly first-order EWPT from the minimal form assumed in the SM.
For example, the EWBG scenario was discussed in models with additional isospin singlets\cite{Espinosa:2011eu,Cline:2012hg,Grzadkowski:2018nbc}, doublets\cite{Turok:1990zg,Cline:1995dg,Fromme:2006cm,Cline:2011mm,Shu:2013uua,Fuyuto:2017ewj,Modak:2018csw} and triplets\cite{Patel:2012pi,Chiang:2014hia}.

One of the important properties of models for the successful EWBG is the strongly first-order phase transition, whose phenomenological consequence can be a prediction on the large deviation in the triple Higgs boson coupling from the SM value\cite{Grojean:2004xa}, in particular, in models with multi Higgs doublets\cite{Kanemura:2004ch}.
Therefore, measuring the triple Higgs boson coupling at future collider experiments, such as the high-luminosity upgrade of the LHC (HL-LHC)\cite{Cepeda:2019klc}, the the International Linear Collider (ILC)\cite{Fujii:2015jha,Fujii:2017vwa}, etc., is important not only to explore the dynamics of electroweak symmetry breaking but also to test the scenario of electroweak baryogenesis.
In addition, the first-order phase transition can also be tested by detecting the gravitational waves with a unique spectrum at future space-based gravitational-wave interferometers such as LISA\cite{Audley:2017drz}, DECIGO\cite{Seto:2001qf} and BBO\cite{Corbin:2005ny} as discussed in Refs.~\cite{Grojean:2006bp,Espinosa:2007qk,Espinosa:2008kw,Espinosa:2010hh,Hindmarsh:2015qta,Kakizaki:2015wua,Hashino:2016rvx,Hashino:2018wee,Zhou:2020xqi}.

The second important property is detecting the effect of CP violation in the Higgs sector.
The CP violation in extended Higgs sectors can be explored by the experiments for electric dipole moment (EDM)\cite{Bernreuther:1990jx,Fukuyama:2012np,Leigh:1990kf,BowserChao:1997bb,Jung:2013hka,Abe:2013qla,Cheung:2014oaa,Cheung:2020ugr}.
By using high-energy collider experiments, it can be tested by searching for the deviations from the SM predictions in the couplings of $ZZZ$, $ZWW$\cite{Degrande:2013kka,Grzadkowski:2016lpv} and $hff$\cite{Barroso:2012wz,Brod:2013cka,Aoki:2018zgq}.
Many methods have been studied in Refs.~\cite{Asakawa:2000jy,Christova:2002uja,Chen:2014ona,Chen:2017plj,Bian:2017jpt}.
The CP violation in extended Higgs sectors can also be tested by measuring the angular distribution of the decays of the SM-like Higgs boson and also those of the extra Higgs bosons\cite{Kuhn:1982di,Hagiwara:2012vz,Harnik:2013aja,Jeans:2018anq,BarShalom:1995jb,Hagiwara:2017ban,Niezurawski:2004ga,Bishara:2013vya}.
However, such additional CP-violating phases are strongly constrained by the EDM data, so that it is generally difficult to keep the sufficient amount of CP-violating phases.

In this paper, we discuss models with an extended Higgs sector which has multiple sources of CP-violating phases.
We investigate the possibility that the effects of CP violation on the EDM are destructive each other, while each CP-violating phase can play an important role of EWBG.
We focus on a general two-Higgs-doublet model (THDM)\cite{Lee:1973iz} as a simple example, in which CP-violating phases appear in the Higgs potential and in the Yukawa interaction.
The general THDM, however, introduces dangerous flavor changing neutral currents (FCNCs) mediated by Higgs bosons.
Thus, we impose the so-called Yukawa-alignment where two-Yukawa interaction matrices for each charged fermion are proportional to each other\cite{Pich:2009sp}.
In addition, we impose another alignment for the Higgs potential such that the Higgs boson with the mass of 125 GeV has the same couplings with the SM particles as those of the SM Higgs boson at tree level because of the measurements of the property of the discovered Higgs boson at the LHC\cite{Sirunyan:2018koj,Aad:2019mbh}.

It is found that the dominant contributions from the Barr-Zee (BZ) diagrams\cite{Barr:1990vd} to the EDM are destructive between fermion-loops and additional scalar boson-loops by keeping $\mathcal{O}(1)$ CP-violating phases and therefore they can be significantly suppressed by the cancellation\footnote{Recently, in Ref.~\cite{Fuyuto:2019svr} another scenario based on the cancellation of the BZ diagrams has been discussed in the general THDM without the alignments in the Yukawa interaction and the Higgs potential. In that paper, the cancellation occurs between the diagrams of fermion-loops and gauge boson-loops.}.
Even when the EDM data can be satisfied by the consequence of the destructive cancellation, each CP-violating phase is not small, so that the model can be tested at future collider experiments by measuring angular distributions of particles generated from the decays of additional Higgs bosons.
Furthermore, we confirm that the above scenario can be perturbatively stable up to a scale much higher than the TeV scale by the analysis using renormalization group equations (RGEs).

This paper is organized as follows.
In Sec.~\ref{sc:model}, we give a brief review of the THDM with CP violation in the Higgs potential and the Yukawa interaction.
Then, in Sec.~\ref{sc:EDM}, we discuss the current constraints from electron and neutron EDMs.
We then explain the destructive interference between the BZ contributions.
In Sec.~\ref{sc:collider}, we consider the collider phenomenology to test the CP-violating phases in our model.
In Sec.~\ref{sc:RGE}, we discuss the scale dependence of the parameters in our scenario by the RGEs.
Finally, we summarize our results in Sec.~\ref{sc:summary}.

%%%%%%%%%%     model     %%%%%%%%%%
\section{Two Higgs doublet model}\label{sc:model}
We consider the general THDM in which the scalar sector consists of two isospin doublets $\Phi_{1,2}$ and the other particle content is the same as the SM.
In this model, the scalar potential $\mathcal{V}$ and the Yukawa interaction term $\mathcal{L}_\textrm{yukawa}$ contain additional CP-violating phases.

%%%%%     potential     %%%%%
\subsection{Higgs potential}
In any basis for the two isodoublet fields $(\primed{\Phi}_1,\primed{\Phi}_2)$, the general Higgs potential is given by
	%%%   potential
		\begin{align}
			\mathcal{V}=
			&	-\primed{\mu}_1^2 (\primed{\Phi}_1^\dagger\primed{\Phi}_1)
				-\primed{\mu}_2^2 (\primed{\Phi}_2^\dagger\primed{\Phi}_2)
				-\sbra{ \primed{\mu}_3^2 (\primed{\Phi}_1^\dagger\primed{\Phi}_2)+h.c.}
			\nn\\
			&	+\tfrac{1}{2} \primed{\lam}_1 (\primed{\Phi}_1^\dagger\primed{\Phi}_1)^2
				+\tfrac{1}{2} \primed{\lam}_2 (\primed{\Phi}_2^\dagger\primed{\Phi}_2)^2
				+\primed{\lam}_3 (\primed{\Phi}_1^\dagger\primed{\Phi}_1) (\primed{\Phi}_2^\dagger\primed{\Phi}_2)
				+\primed{\lam}_4 (\primed{\Phi}_2^\dagger\primed{\Phi}_1) (\primed{\Phi}_1^\dagger\primed{\Phi}_2)
			\nn\\
			&	+\cbra{\sbra{
					\tfrac{1}{2} \primed{\lam}_5 (\primed{\Phi}_1^\dagger\primed{\Phi}_2)
					+\primed{\lam}_6 (\primed{\Phi}_1^\dagger\primed{\Phi}_1)
					+\primed{\lam}_7 (\primed{\Phi}_2^\dagger\primed{\Phi}_2)
				} (\primed{\Phi}_1^\dagger\primed{\Phi}_2)+h.c.}
		\label{eq:potential1}
		,\end{align}
where $\primed{\mu}_{1,2}^2$ and $\primed{\lam}_{1,2,3,4}$ are real, while $\primed{\mu}_3^2$ and $\primed{\lam}_{5,6,7}$ are complex.
The potential is unchanged under the U(2) transformation between the doublet fields.
When the neutral components of the doublet fields get the vacuum expectation values (VEVs), $(\eval{\primed{\Phi}_1^0},\eval{\primed{\Phi}_2^0})=(v_1 e^{i\xi_1}/\sqrt{2},v_2 e^{i\xi_2}/\sqrt{2})$, so-called the Higgs basis\cite{Botella:1994cs,Davidson:2005cw} can be realized by the U(2) transformation as
	%%%   basis transformation
		\begin{align}
			\rbra{\begin{array}{c}
				\Phi_1 \\ \Phi_2
			\end{array}}
			=
			\rbra{\begin{array}{rr}
				\cos\beta &\sin\beta \\-\sin\beta & \cos\beta
			\end{array}}
			\rbra{\begin{array}{rr}
				e^{-i\xi_1} &0 \\0 & e^{-i\xi_2}
			\end{array}}
			\rbra{\begin{array}{c}
				\primed{\Phi}_1 \\ \primed{\Phi}_2
			\end{array}},
		\label{eq:basistransformation}
		\end{align}
where $\tan\beta\equiv v_2/v_1$.
The two doublet fields $\Phi_1$ and $\Phi_2$ are parametrized as
	%%%   scalar doublets
		\begin{align}
			\Phi_1=\rbra{\begin{array}{c} G^+\\\frac{1}{\sqrt{2}}(v+h^0_1+iG^0) \end{array}}
			,\quad
			\Phi_2=\rbra{\begin{array}{c} H^+\\\frac{1}{\sqrt{2}}(h^0_2+ih^0_3) \end{array}}
		\label{eq:parametrizeddoublets}
		,\end{align}
where $G^+$ and $G^0$ are the Nambu-Goldstone bosons and $v\equiv\sqrt{v_1^2+v_2^2}=(\sqrt{2}G_F)^{-1/2}$ with $G_F$ being the Fermi constant.
In the Higgs basis, the potential is rewritten by
	%%%   potential
		\begin{align}
			\mathcal{V}=
			&	-\mu_1^2 (\Phi_1^\dagger\Phi_1)
				-\mu_2^2 (\Phi_2^\dagger\Phi_2)
				-\sbra{ \mu_3^2 (\Phi_1^\dagger\Phi_2)+h.c.}
			\nn\\
			&	+\tfrac{1}{2} \lam_1 (\Phi_1^\dagger\Phi_1)^2
				+\tfrac{1}{2} \lam_2 (\Phi_2^\dagger\Phi_2)^2
				+\lam_3 (\Phi_1^\dagger\Phi_1) (\Phi_2^\dagger\Phi_2)
				+\lam_4 (\Phi_2^\dagger\Phi_1) (\Phi_1^\dagger\Phi_2)
			\nn\\
			&	+\cbra{\sbra{
					\tfrac{1}{2} \lam_5 (\Phi_1^\dagger\Phi_2)
					+\lam_6 (\Phi_1^\dagger\Phi_1)
					+\lam_7 (\Phi_2^\dagger\Phi_2)
				} (\Phi_1^\dagger\Phi_2)+h.c.}
		\label{eq:potential2}
		.\end{align}
These parameters, $\mu_i^2$ and $\lam_i$, can be expressed by $\primed{\mu}_i^2$ and $\primed{\lam}_i$ (see Appendix~\ref{sc:parameters}), and $\mu_{1,2}^2$ and $\lam_{1,2,3,4}$ are real, while $\mu_3^2$ and $\lam_{5,6,7}$ are complex.
The stationary conditions
	%%%   stationary condition
		\begin{align}
			0=	\left.
					\frac{\partial\mathcal{V}}{\partial h^0_j}
				\right|_{\substack{\Phi_1=\eval{\Phi_1}\\\Phi_2=\eval{\Phi_2}}}
		,\end{align}
lead to
		\begin{align}
			\mu_1^2 = \frac{1}{2}\lam_1v^2
		,~	\mu_3^2 = \frac{1}{2}\lam_6v^2
		.\end{align}
The remaining demensionful parameter is redefined as $M^2\equiv-\mu_2^2$.
The squared mass of the charged Higgs boson is given by
	%%%   charged higgs mass
		\begin{align}
			m_{H^\pm}^2=M^2+\tfrac{1}{2}\lam_3v^2
		.\end{align}
The squared-mass matrix for the neutral Higgs bosons in the basis of $(h^0_1,h^0_2,h^0_3)$ is given by
	%%%   mass matrix for h^0
		\small
		\begin{align}
			\mathcal{M}^2_{ij}
				\equiv	\left.
						\frac{\partial^2\mathcal{V}}{\partial h_i\partial h_j}
					\right|_{\substack{\Phi_1=\eval{\Phi_1}\\\Phi_2=\eval{\Phi_2}}}
				=	v^2
					\rbra{
					\begin{array}{ccc}
						\lam_1	&\re[\lam_6]	&-\im[\lam_6]	\\
						\re[\lam_6]	&\frac{M^2}{v^2}+\frac{1}{2}(\lam_3+\lam_4+\re[\lam_5])	&-\frac{1}{2}\im[\lam_5]	\\
						-\im[\lam_6]	&-\frac{1}{2}\im[\lam_5]	&\frac{M^2}{v^2}+\frac{1}{2}(\lam_3+\lam_4-\re[\lam_5])
					\end{array}
					}_{ij}
		\label{eq:massmatrix}
		.\end{align}
		\normalsize
This is diagonalized as $\mathcal{R}^T\mathcal{M}^2\mathcal{R}=\mathrm{diag}(m_{H^0_1}^2,m_{H^0_2}^2,m_{H^0_3}^2)$, where $\mathcal{R}$ is the orthogonal matrix.
The mass eigenstates of the neutral scalars are expressed as
		\eq{
			h^0_i=\mathcal{R}_{ij}H^0_j
		.}

In the model, one of the phase of parameters can be absorbed by redefinition of the doublet fields.
Consequently, the potential has 11 parameters: $v, M, \lam_{1,2,3,4}, |\lam_{5,6,7}|$ and the 2 physical phases.
Hereafter, we take $\im[\lam_5]=0$ by using the phase redefinition,
	%%%   phase redefinition
		\texteq{
			(\Phi_1^\dagger\Phi_2)\to e^{-\arg[\lam_5]/2}(\Phi_1^\dagger\Phi_2)
		,}
and we also redefine the other complex parameters as
		\texteq{
			\mu_3^2e^{-\arg[\lam_5]/2}\to\mu_3^2,
			\lam_6e^{-\arg[\lam_5]/2}\to\lam_6
			~\textrm{and}~
			\lam_7e^{-\arg[\lam_5]/2}\to\lam_7
		.}

We impose the condition of the potential alignment,
	%%%   potential alignment
		\eq{
			\lam_6=0
		,}
in Which $\mathcal{R}_{ij}=\delta_{ij}$, so that the neutral scalar bosons do not mix with each other. The squared masses of the neutral scalars are then given by
	%%%   mass of neutral scalars
		\eq{
			m_{H^0_1}^2 &= \lam_1 v^2
		,\\	m_{H^0_2}^2 &= M^2+\frac{1}{2}(\lam_3+\lam_4+\re[\lam_5]) v^2
		,\\	m_{H^0_3}^2 &= M^2+\frac{1}{2}(\lam_3+\lam_4-\re[\lam_5]) v^2
		.}
We identify $H^0_1$ as the discovered Higgs boson with the mass of 125 GeV, and we consider that the other Higgs bosons $H^0_2, H^0_3$ and $H^\pm$ have larger masses.
Consequently, there are 7 free parameters which can be chosen as follows
	%%%   parameters
		\eq{
			m_{H^0_2}, m_{H^0_3}, m_{H^\pm}, M, \lam_2, |\lam_7| ~\textrm{and}~ \theta_7,
		}
where $\theta_7\equiv\arg[\lam_7]\in(-\pi,\pi]$.

%%%%%     yukawa     %%%%%
\subsection{Yukawa interaction}
The Yukawa interaction term is given by
	%%%   yukawa term
		\begin{align}
			\mathcal{L}_\textrm{yukawa}=
				-\sum_{k=1}^2
					\rbra{
					\bar{\primed{Q}}_L \primed{y}_{u,k}^{\dagger} \tilde{\primed{\Phi}}_k \primed{u}_R
					+\bar{\primed{Q}}_L \primed{y}_{d,k} \primed{\Phi}_k \primed{d}_R
					+\bar{\primed{L}}_L \primed{y}_{e,k} \primed{\Phi}_k \primed{e}_R
					+h.c. },
		\label{eq:yukawa1}
		\end{align}
where $\primed{Q}_L$ ($\primed{L}_L$) is the left-handed quark (lepton) doublet, and $\primed{u}_R$, $\primed{d}_R$ and $\primed{e}_R$ are the right-handed up-type quark, down-type quark and charged lepton singlets.
In Eq.~\eqref{eq:yukawa1}, $\tilde{\primed{\Phi}}_{k}=i \sigma_2 \primed{\Phi}_k^*$, and $\primed{y}_{f,k}$ ($f=u$, $d$ and $e$) are $3\times3$ complex Yukawa coupling matrices.
The Yukawa interaction term can be also expressed in the Higgs basis as follows
	%%%   yukawa term
		\begin{align}
			\mathcal{L}_\textrm{yukawa}=
				-\sum_{k=1}^2
					\rbra{
					\bar{\primed{Q}}_L y_{u,k}^{\dagger} \tilde{\Phi}_k \primed{u}_R
					+\bar{\primed{Q}}_L y_{d,k} \Phi_k \primed{d}_R
					+\bar{\primed{L}}_L y_{e,k} \Phi_k \primed{e}_R
					+h.c. },
		\label{eq:yukawa2}
		\end{align}
where the Yukawa coupling matrices are expressed as
	%%%   yukawa matrix
		\begin{align}
			y_{f,1}&=\rbra{\primed{y}_{f,1}e^{i\xi_1}}\cos\beta+\rbra{\primed{y}_{f,2}e^{i\xi_2}}\sin\beta
		,\\	y_{f,2}&=-\rbra{\primed{y}_{f,1}e^{i\xi_1}}\sin\beta+\rbra{\primed{y}_{f,2}e^{i\xi_2}}\cos\beta
		.\end{align}
In the mass eigenstate of the fermions represented as those without a prime, Eq.~\eqref{eq:yukawa2} is rewritten as
	%%%   yukawa term
		\begin{align}
			\mathcal{L}_\textrm{yukawa}=
				&	-\bar{Q}_L^u \rbra{
						\sqrt{2}\frac{M_u}{v}\tilde{\Phi}_1 +\rho_u\tilde{\Phi}_2
					} u_R
					-\bar{Q}_L^d \rbra{
						\sqrt{2}\frac{M_d}{v}\Phi_1 +\rho_d\Phi_2
					}d_R
				\nn\\
				&	-\bar{L}_L \rbra{
						\sqrt{2}\frac{M_e}{v}\Phi_1 +\rho_e\Phi_2
					} e_R
					+h.c.,
		\end{align}
where $Q_L^u=(u_L,V_\textrm{CKM}d_L)^T$ and $Q_L^d=(V_\textrm{CKM}^\dagger u_L,d_L)^T$ with $V_\textrm{CKM}$ being the Cabibbo-Kobayashi-Maskawa (CKM) matrix.
In the above expression, $M_f$ are the diagonalized-mass matrices for the fermions and $\rho_f$ are the general complex $3\times3$ matrices.
The off-diagonal elements of $\rho_f$ generally induce FCNCs which are strongly constrained by flavor experiments.

In order to avoid such FCNCs, we impose the Yukawa-alignment proposed by Pich and Tuzon\cite{Pich:2009sp} as
	%%%   yukawa alignment
		\eq{
			y_{f,2}=\zeta_f ~y_{f,1}
		,}
where $\zeta_f$ are complex values.
Thus, $y_{f,1}$ and $y_{f,2}$ are diagonalized at the same time, and then $\rho_f$ is expressed by
	%%%   extra yukawa matrix
		\texteq{
			\rho_u=\frac{\sqrt{2}}{v}M_u\zeta_u^*
		} and
		\texteq{	
			\rho_{d,e}=\frac{\sqrt{2}}{v}M_{d,e}\zeta_{d,e}
		.}
There is another prescription suggested by Glashow and Weinberg\cite{Glashow:1976nt}, in which a $Z_2$ symmetry is imposed to the Higgs sector.
The doublet fields are transformed as $\primed{\Phi}_1\to\primed{\Phi}_1$ and $\primed{\Phi}_2\to-\primed{\Phi}_2$ under the $Z_2$ symmetry.
One of the Yukawa matrices $\primed{y}_{k,f}$ is then forbidden.
There are four types of Yukawa interactions depending on the $Z_2$ charge assignment for the right-handed fermions\cite{Barger:1989fj,Aoki:2009ha}.
The $\zeta_f$ factors in each model are summarized in Tab.~\ref{tb:zeta}.
	%%%   tab: zeta
		\begin{table}
			\centering
			\begin{tabular}{|c||c|c|c|} \hline
				Model			& $\zeta_u$ &$\zeta_d$ & $\zeta_l$ \\ \hline \hline
				Our model		& arbitrary complex & arbitrary complex & arbitrary complex \\ \hline
				Type-I THDM		& $1/\tan\beta$ & $1/\tan\beta$ & $1/\tan\beta$ \\
				Type-II THDM		& $1/\tan\beta$ & $-\tan\beta$ & $-\tan\beta$ \\
				Type-X THDM	& $1/\tan\beta$ & $1/\tan\beta$ & $-\tan\beta$ \\
				Type-Y THDM		& $1/\tan\beta$ & $-\tan\beta$ & $1/\tan\beta$ \\ \hline
			\end{tabular}
			\caption{$\zeta_f$ factors in the THDMs. The names of the models are referred to Refs.~\cite{Pich:2009sp,Aoki:2009ha}.}
			\label{tb:zeta}
		\end{table}

The Yukawa interaction term can be written by the mass eigenstates of the fermions and the Higgs bosons as follows:
	%%%   yukawa term in mass basis
		\begin{align}
			\mathcal{L}_\textrm{yukawa} \supset
					&	-\sum_{f=u,d,e}
						\cbra{
						\bar{f}_L M_f f_R
						+\sum_{j=1}^3\bar{f}_L \rbra{\frac{M_f}{v} \kappa_f^j} f_R H^0_j+h.c.
						}
					\nn\\
					&	-\frac{\sqrt{2}}{v}
						\cbra{
							-\zeta_u \bar{u}_R (M_u^\dagger V_\textrm{CKM}) d_L
							+\zeta_d \bar{u}_L (V_\textrm{CKM} M_d) d_R
							+\zeta_e \bar{\nu}_L M_e e_R
						} H^{+}+h.c.
		.\end{align}
We introduce $\kappa_f^j$ as the coupling factors for the interactions of the neutral Higgs bosons and fermions defined as
	%%%   kappa
		\begin{align}
			\kappa_f^j=\mathcal{R}_{1j}
						+\sbra{
							\mathcal{R}_{2j} +i(-2I_f)\mathcal{R}_{3j}
						} |\zeta_f|e^{i(-2I_f)\theta_f}
		,\end{align}
where $\theta_f\equiv\arg[\zeta_f]\in(-\pi,\pi]$ and $I_u=1/2, I_d=I_e=-1/2$.
As we mentioned above, $\mathcal{R}_{jk}=\delta_{jk}$ is taken, so that $\kappa$ factors are written as
	%%%   kappa
		\begin{align}
			\kappa_f^{1} &=1
		,\\	\kappa_f^{2} &=|\zeta_f| e^{i(-2I_f)\theta_f}
		,\\	\kappa_f^{3} &=i(-2I_f) \kappa_f^2
		.\end{align}
We can see that the Yukawa couplings for $H^0_1$ do not contain the CP-violating phases at tree-level.
On the other hand, those of $H^0_{2,3}$ have the CP-violating phases, and thus they can contribute to the EDMs.

%%%%%     kinetic term   %%%%%
\subsection{Kinetic terms of the scalar fields}
The kinetic term for the scalar doublet fields can be rewritten by
	%%%   kinetic term
		\begin{align}
			\mathcal{L}_\textrm{kin}
				=|D_\mu\primed{\Phi}_1|^2+|D_\mu\primed{\Phi}_2|^2
				=|D_\mu\Phi_1|^2+|D_\mu\Phi_2|^2,
		\end{align}
where $D_\mu$ is the covariant derivative given by
	%%%   covariant derivative for doublets
		\texteq{
			D_\mu=\partial_\mu+i g_2 \frac{\sigma^a}{2} W_\mu^a +i g_1 \frac{1}{2} B_\mu
		}
with $g_2$ and $g_1$ being the SU(2)$_L$ and U(1)$_Y$ gauge coupling constants, respectively.
The trilinear Higgs-gauge-gauge type couplings are given by
	%%%   HVV couplings
		\begin{align}
			\mathcal{L}_\textrm{kin}\supset
				\sum_j^3\mathcal{R}_{1j}
					\rbra{
						\frac{2m_W^2}{v}W_\mu W^\mu
						+\frac{m_Z^2}{v}Z_\mu Z^\mu
					} H^0_j
		,\end{align}
where $m_W=g_2v/2$ and $m_Z=\sqrt{g_2^2+g_1^2}~v/2$.
In the alignment limit $\mathcal{R}_{jk}=\delta_{jk}$, the couplings of $H^0_1VV$ ($V=W,~Z$) are the same as the SM ones, and those of $H^0_2VV$ and $H^0_3VV$ vanish at tree level.

%%%%%     theoretical and experimental constraints   %%%%%
\subsection{Theoretical and experimental constraints}
The parameters in the potential are constrained by taking into account the perturbative unitarity\cite{Ginzburg:2005dt,Kanemura:1993hm,Akeroyd:2000wc,Kanemura:2015ska} and the vacuum stability\cite{Nie:1998yn,Kanemura:1999xf}.
In addition, the electroweak $S$, $T$ and $U$ parameters\cite{Peskin:1990zt,Peskin:1991sw} constrain the masses and mixings of the Higgs bosons.
When we impose $m_{H_3}=m_{H^\pm}$ and $\lam_6=0$, new contributions to the $T$ parameter vanish at one-loop level, 
because of the custodial symmetry in the Higgs potential\cite{Pomarol:1993mu,Haber:1992py,Grzadkowski:2010dj,Haber:2010bw,Kanemura:2011sj}.
Furthermore, the constraints from $B$ physics should be considered\cite{Mahmoudi:2009zx}.
The constraint on the mass of charged Higgs bosons and $\zeta_q$ in the Yukawa-alignment scenario has been discussed in Ref.~\cite{Jung:2010ab}.

%%%%%%%%%%     EDM     %%%%%%%%%%
\section{Electric dipole moment}\label{sc:EDM}
The Hamiltonian of the EDM for non-relativistic particles with the spin $\vec{S}$ can be described by
	%%%   EDM Hamiltonian
		\eq{
			H_\textrm{EDM}=-d_f \frac{\vec{S}}{|\vec{S}|}\cdot\vec{E}
		,}
where $\vec{E}$ is the external electric field.
Under the time reversal transformation $\mathcal{T}(\vec{S})=-\vec{S}$ and $\mathcal{T}(\vec{E})=+\vec{E}$, the sign of this term $H_\textrm{EDM}$ is flipped.
Therefore, if the EDM is nonzero, time reversal invariance is broken, then the CPT theorem implies that CP symmetry is also broken.
In terms of the effective Lagrangian, the EDM $d_f$ for a fermion is written as
	%%%   EDM Lagrangian
		\begin{align}
			\mathcal{L}_\textrm{EDM}=-\frac{d_f}{2}\bar{f} \sigma^{\mu\nu}(i\gamma^5)f F_{\mu\nu}
		,\end{align}
where $F_{\mu\nu}$ is the electromagnetic field strength tensor and
	%%% sigma mu nu
		\texteq{
			\sigma^{\mu\nu}=\frac{i}{2}[\gamma^\mu,\gamma^\nu]
		.}
The neutron EDM receives the additional contribution from the chromo-EDM (CEDM) $d_q^C$ for a quark $q$ being expressed as
	%%%   CEDM Lagrangian
		\begin{align}
			\mathcal{L}_\textrm{CEDM}=-\frac{d_q^C}{2}\bar{q} \sigma^{\mu\nu}(i\gamma^5)q G_{\mu\nu}
		,\end{align}
where the $G_{\mu\nu}$ is the QCD field strength tensor.

The electron and neutron EDMs are constrained by various atomic and molecule experiments\cite{Jung:2013hka,Cheung:2014oaa} as follows.
First, the ACME collaboration gave the upper limit $|d_e+kC_S|<1.1\times10^{-29}$ e~cm at the 90\% confidence level (CL)\cite{Andreev:2018ayy} by using the thorium-monoxide EDM, where $C_S$ is defined by the coefficient of the following dimension six operator
	%%%   (e gamma5 e)(NN)
		\eq{
			\mathcal{L}\supset C_S(\bar{e}i\gamma_5 e)(\bar{N}N)
		,}
with $N$ being a nucleon.
The coefficient $k$ is given by $k\sim \mathcal{O}(10^{-15})$ GeV${}^2$ e~cm\cite{Cheung:2014oaa}.
In our benchmark scenario, which is explained below, the contribution from $kC_S$ is typically two orders of the magnitude smaller than the current bound, according to the discussion given in Refs.~\cite{Cheung:2014oaa,Jung:2013hka}.
Therefore, we can safely neglect this contribution, and we simply impose the bound $|d_e|<1.1\times10^{-29}$ e~cm (90\% CL) in the following discussion.

Second, very recently the nEDM collaboration updated the constraint on the neutron EDM as $|d_n|<1.8\times10^{-26}$ e~cm (90\% CL)\cite{Abel:2020gbr}.
In our analysis, the neutron EDM is estimated by using the following QCD sum rule\cite{Abe:2013qla}:
	%%%   nEDM
		\begin{align}
			d_n=0.79d_d-0.20d_u+e(0.59d_d^C+0.30d_u^C)/g_3
		\label{eq:nEDM}
		,\end{align}
where $g_3$ is the QCD gauge coupling constant.
There are the other contributions to the neutron EDM from the Weinberg operator $\mathcal{L}\supset\frac{1}{3}C_WG^a_{\mu\nu}\tilde{G}^b{}^{\nu\sigma}G^c{}_{\sigma}{}^{\mu}$ and the four-fermi interaction $\mathcal{L}\supset C_{ff'}(\bar{f}f)(\bar{f}'i\gamma_5f')$.
However, in our benchmark scenario, the contribution from $C_W (C_{ff'})$ is typically two (more than two) orders of the magnitude smaller than the current bound, according to the discussion given in Refs.~\cite{Bernreuther:1990jx,Cheung:2014oaa,Jung:2013hka} (Ref.~\cite{Jung:2013hka}).
We thus simply apply Eq.~\eqref{eq:nEDM} to our analysis.

Finally, the bounds from the other EDMs of atoms are satisfied by considering the constraints for the above electron and the neutron EDMs.
Consequently, what we need to take into account is the contributions from $d_f$ ($f=e$, $d$ and $u$) and $d^C_q$ ($q=d$ and $u$).

The dominant contribution to $d_f$ appears from the two-loop BZ type diagrams\cite{Barr:1990vd}.
We note that the one-loop contributions to $d_f$ are negligibly smaller than those from the BZ diagrams, because the one-loop contribution has two additional powers of small Yukawa couplings for light fermions with respect to the BZ diagrams.
In fact, we obtain $d_e(\textrm{1-loop})\sim \mathcal{O}(10^{-34})$ e~cm, while $d_e(\textrm{2-loop})\sim \mathcal{O}(10^{-27})$ e~cm in the typical input parameters with $\mathcal{O}(100)$ GeV of the masses of additional Higgs bosons and the $\kappa_f$ factors of $\mathcal{O}(1)$\cite{Bernreuther:1990jx}.

\begin{figure}
			\begin{tabular}{c}
				%%%
				\begin{minipage}{0.5\hsize}
					\centering
					\includegraphics[width=50mm]{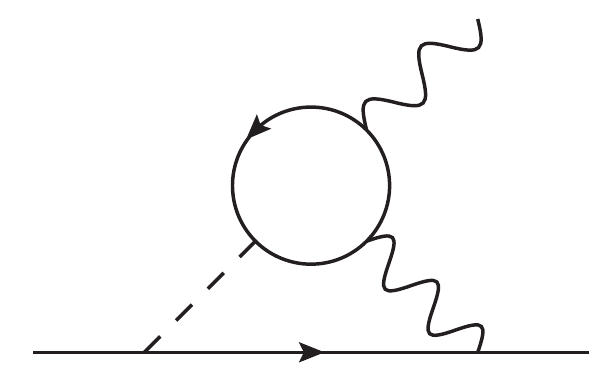}
					\\(\ref{fg:BZ}-1) Fermion-loop
				\end{minipage}
				%%%
				\begin{minipage}{0.5\hsize}
					\centering
					\includegraphics[width=50mm]{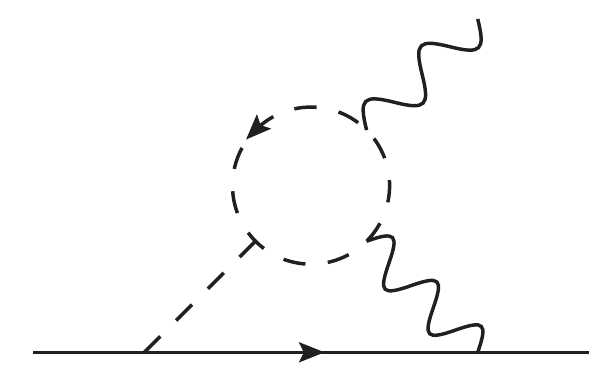}
					\\(\ref{fg:BZ}-2) Higgs boson-loop
				\end{minipage}
				%%%
			\end{tabular}
			\caption{BZ type diagrams contributing to the electron EDM.}
			\label{fg:BZ}
		\end{figure}

The BZ diagrams contain the effective $H^0_jV^0\gamma$ ($V^0=\gamma, Z$) and $H^\pm W^\mp\gamma$ vertices.
The BZ contribution to $d_f$ is decomposed into the contributions from fermion-loops, Higgs boson-loops and gauge bosom-loops as follows
	%%%   BZ EDM
		\eq{
			d_f=d_f(\textrm{fermion})+d_f(\textrm{Higgs})+d_f(\textrm{gauge})
		\label{eq:BZcontribution1}
		.}
Furthermore, each contribution can be classified as
		\eq{
			d_f(X)=d_f^\gamma(X)+d_f^Z(X)+d_f^W(X)
		\label{eq:BZcontribution2}
		,}
where $d_f^\gamma(X)$, $d_f^Z(X)$ and $d_f^W(X)$ are the contributions from diagrams with $\gamma$, $Z$ and $W$ exchange, respectively.
The explicit expressions for each component of $d_f$ are given in Appendix~\ref{sc:BZformulae} in the Yukawa-aligned THDM.
The gauge boson-loop contributions $d_e(\textrm{gauge})$ are proportional to $\sum_{j=1}^3\mathcal{R}_{1j}\im[\kappa_f^j]$\footnote{When we take all the neutral Higgs bosons degenerate in mass, $d_e$ also vanishes because of the orthogonality of the $\mathcal{R}$ matrix\cite{Jung:2013hka}.}.
Thus, by imposing the potential alignment $\mathcal{R}_{ij}=\delta_{ij}$ this contribution vanishes.
In addition, it is confirmed in Refs.\cite{Leigh:1990kf,Jung:2013hka} that non-BZ type diagrams at two-loop level with a Higgs boson mediation are also proportional to $\sum_{j=1}^3\mathcal{R}_{1j}\im[\kappa_f^j]$, so that they vanish in the alignment limit.
Therefore, the dominant contributions arise from the Higgs boson and fermion-loops as shown in Fig.~\ref{fg:BZ}.
	%%%   fig: BZ diagram

Requiring the destructive interference between the fermion-loop and Higgs boson-loop in order to realize $d_e\simeq0$, we obtain the following relation assuming the extra Higgs boson masses to be nearly degenerate and $\theta_d=0$,
	%%%   destructive EDM
		\begin{align}
			\rbra{A^\gamma I^\gamma+A^Z I^Z+A^W I^W}|\zeta_u| \sin(\theta_u-\theta_e)
			\simeq
			-\rbra{B^\gamma J^\gamma+B^Z J^Z+B^W J^W}|\lam_7| \sin(\theta_7-\theta_e)
		\label{eq:destructiveEDM}
		,\end{align}
where $A$ and $B$ are the constant factors:
	%%%
		\begin{align}
				A^\gamma	&=\frac{32}{3}\sin^2\theta_W\frac{m_t^2}{v^2}
			,\\	A^Z			&=\frac{1}{3}\frac{(1-4\sin^2\theta_W) (3-8\sin^2\theta_W)}{\cos^2\theta_W}\frac{m_t^2}{v^2}
			,\\	A^W		&=3\frac{m_t^2}{v^2}
			,\\	B^\gamma	&=4\sin^2\theta_W
			,\\	B^Z			&=\frac{1}{2}\frac{(1-4\sin^2\theta_W)}{\cos^2\theta_W} \cos{2\theta_W}
			,\\	B^W			&=-\frac{1}{2}
		,\end{align}
and $I$ and $J$ are the loop functions depending on the masses of the extra Higgs bosons:
	%%%
		\begin{align}
				I^V	&=2\int_0^1dz \sbra{\frac{1}{z}-(1-z)} C^{V\tilde{H}}_{tt}(z)
				,\quad\quad(V=\gamma~\textrm{and}~Z)
			,\\	I^W	&=\int_0^1dz \frac{2-z}{z}\sbra{\frac{2}{3}-z} C^{W\tilde{H}}_{tb}(z)
			,\\	J^V	&=2\int_0^1dz (1-z) C^{V\tilde{H}}_{\tilde{H}\tilde{H}}(z)
				,\quad\quad\quad(V=\gamma,Z~\textrm{and}~W)
		,\end{align}
where $C^{GH}_{XY}(z)$ are given in Appendix~\ref{sc:BZformulae}, and its argument $m_{\tilde{H}}$ denotes the mass of the extra Higgs bosons in the loop.
From Eq.~\eqref{eq:destructiveEDM}, it is clear that the two independent phases $\theta_u$ and $\theta_7$ can be taken such that the fermion- and the Higgs boson-loop contributions to $d_e$ cancel with each other.

\begin{figure}
			\centering
			\includegraphics[width=80 mm]{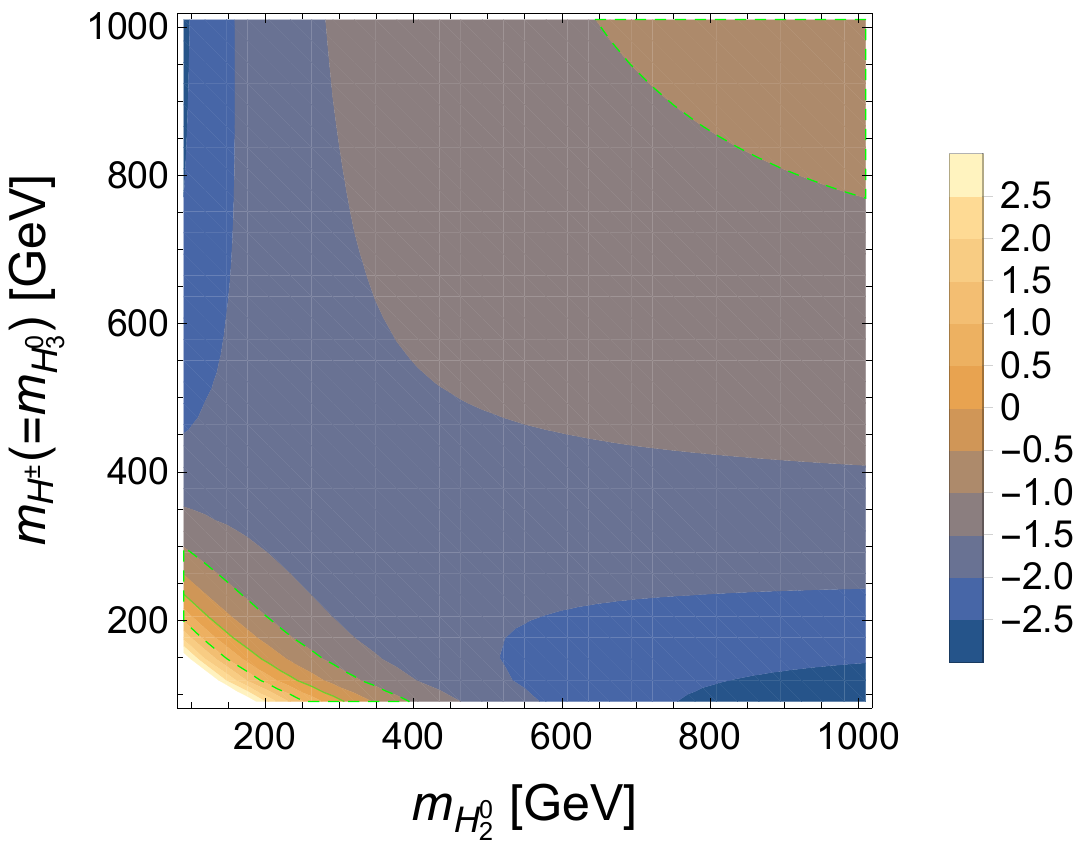}
			\caption{
				Contour plot for the electron EDM $d_e$ in the unit of 10$^{-29}$ e~cm as a function of $m_{H^0_2}$ and $m_{H^\pm}(=m_{H^0_3})$ with $(\theta_d, \theta_e, \theta_u, \theta_7)=(2,0,\pi/2,1)$, $|\lam_7|=0.3$, $|\zeta_u|=0.01$, $|\zeta_d|=0.1$ and $|\zeta_e|=0.5$ at the scale $m_Z$ in the panel.
				The green solid line denotes $d_e=0$ and the area between the green dashed lines is the allowed region against the constraint from the electron EDM.
			}
			\label{fg:massmasscontour}
		\end{figure}

We numerically calculate the electron EDM by using the input parameters in the SM in Tab.~\ref{tb:SMinput}.
	%%%   tab: inputs of SM para
		\begin{table}
			\centering
			\begin{tabular}{lll}
				\hline
					$m_u=1.29\times10^{-3}$,
				&	$m_c=0.619$,
				&	$m_t=171.7$,
				\\
					$m_d=2.93\times10^{-3}$,
				&	$m_s=0.055$,
				&	$m_b=2.89$,
				\\
					$m_e=0.487\times10^{-3}$,
				&	$m_\mu=0.103$,
				&	$m_\tau=1.746$	~~(in GeV).
				\\ \hline
			\end{tabular}
			\\
			\begin{tabular}{cccc}
					$\alpha_\textrm{em}=1/127.955$,
				&	$m_Z=91.1876$ GeV,
				&	$m_W=80.379$ GeV,
				&	$\alpha_S=0.1179$.
				\\ \hline
			\end{tabular}
			\\
			\begin{tabular}{cccc}
					$\lambda=0.22453$,
				&	$A=0.836$,
				&	$\bar{\rho}=0.122$,
				&	$\bar{\eta}=0.355$.
				\\ \hline
			\end{tabular}
			\caption{
				Input values for the SM parameters at the scale of $m_Z$.
				We refer to Refs.~\cite{Xing:2007fb,Bijnens:2011gd} for the fermion masses at $m_Z$.
				The other parameters in the table are taken in Ref.~\cite{Tanabashi:2018oca}.
				The parameters in the last line are the Wolfenstein parameters of the CKM matrix\cite{Tanabashi:2018oca}.
			}
			\label{tb:SMinput}
		\end{table}
As we mentioned in Sec.~\ref{sc:model}, we take $m_{H^0_3}=m_{H^\pm}$ in order to avoid the constraint from the $T$ parameter in the following analysis.
In Fig.~\ref{fg:massmasscontour}, the electron EDM is shown by the contour plot as a function of $m_{H^0_2}$ and $m_{H^\pm}(=m_{H^0_3})$ with $(\theta_u, \theta_d, \theta_e, \theta_7)=(2,0,\pi/2,1)$, $|\lam_7|=0.3$, $|\zeta_u|=0.01$, $|\zeta_d|=0.1$ and $|\zeta_e|=0.5$ at the scale $m_Z$.
The green solid line denotes $d_e=0$ and the area between green dashed lines is the allowed region against the constraint from the electron EDM.
It is seen that $d_e$ is getting close to $0$ when the masses of the additional Higgs bosons are taken to be $\mathcal{O}(200)$ GeV.
If one of the masses is taken to be much larger than that of the others, $d_e$ is getting larger because the cancellation becomes weaker.
If all the masses are simultaneously taken to be larger up to about 400 GeV, $d_e$ becomes larger because of the same reason.
If they are further taken to be larger, $d_e$ turns out to be decreasing due to the decoupling behavior of the additional Higgs bosons.
	%%%   fig: eEDM mass-mass contour

\begin{figure}
			\centering
			\includegraphics[height=70 mm]{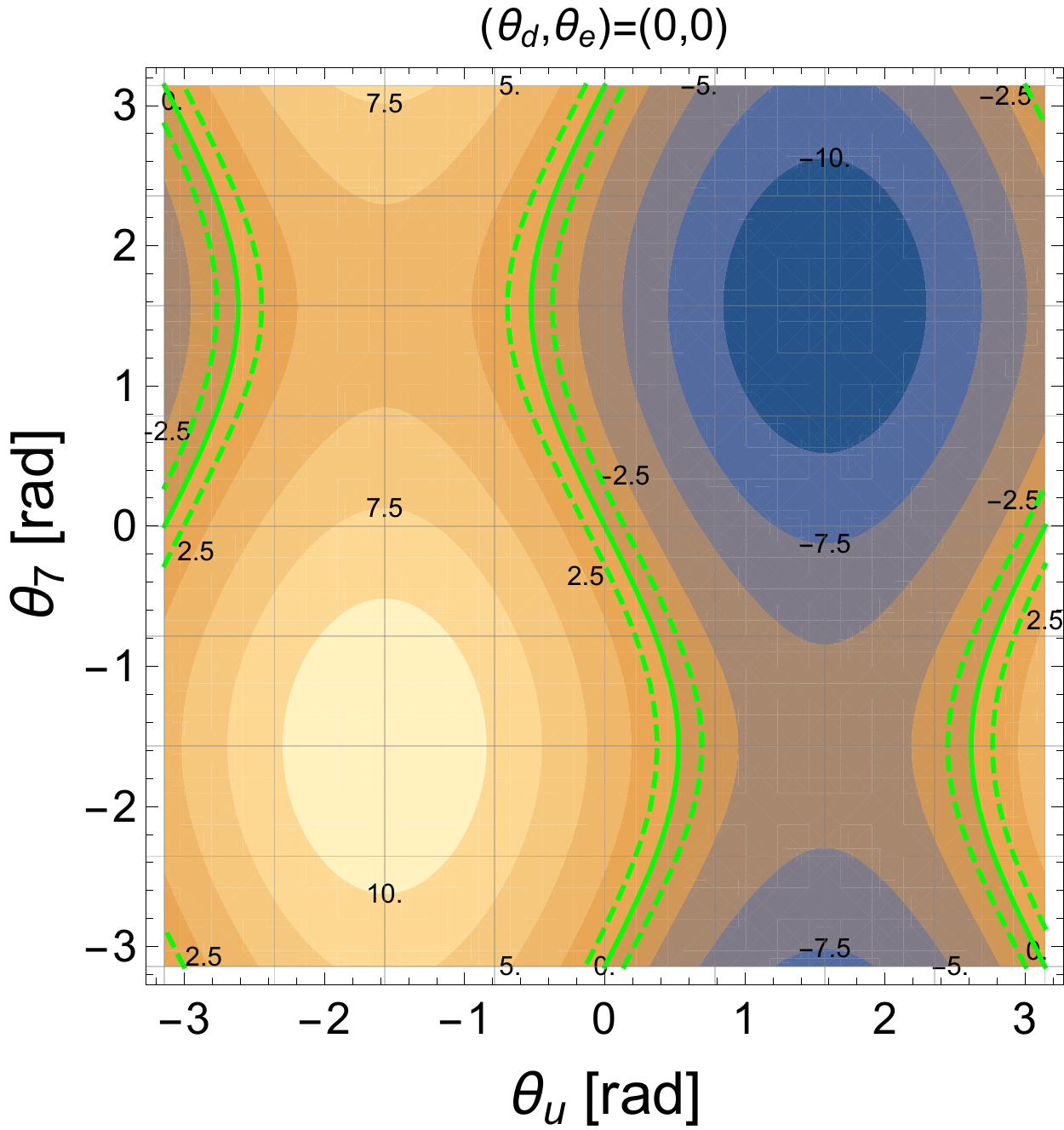}
			\includegraphics[height=70.5 mm]{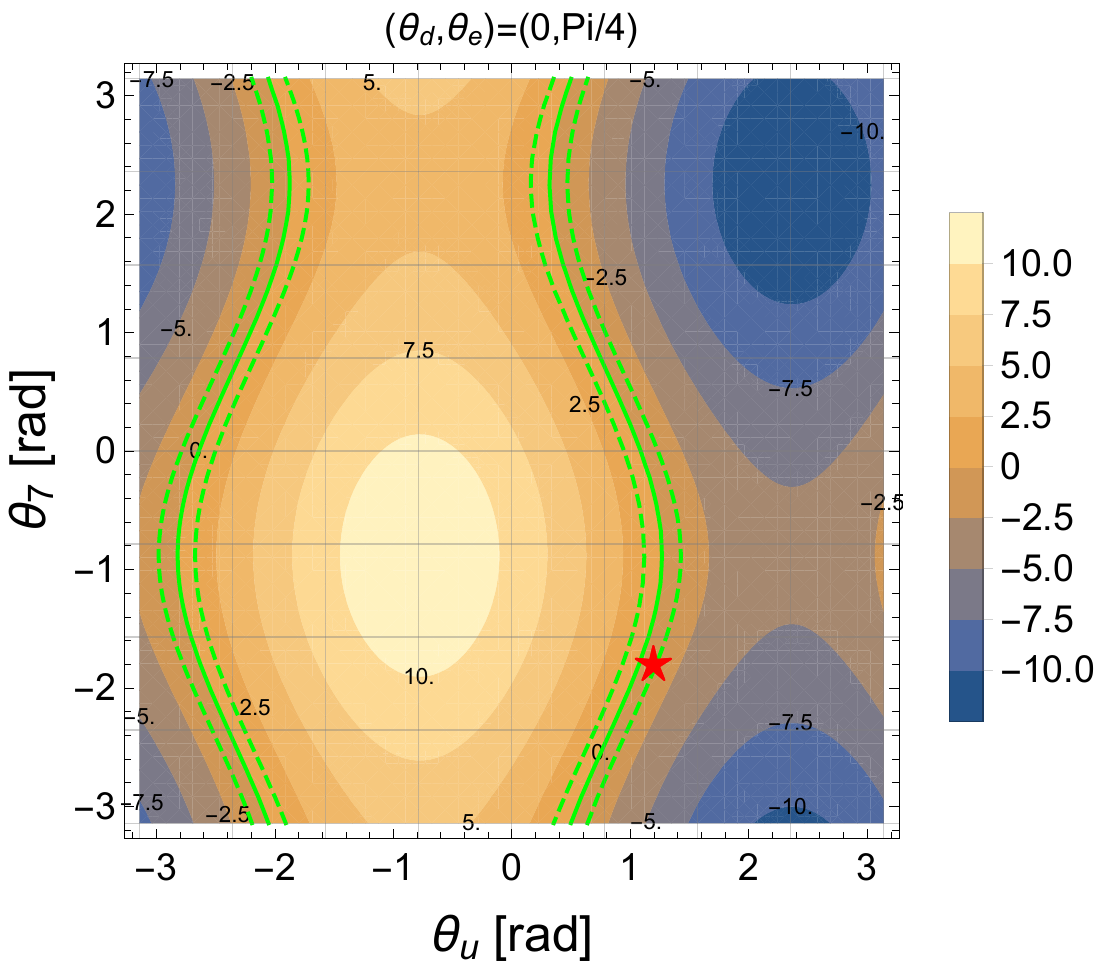}
			\caption{
				Contour plot for the electron EDM $d_e$ in the unit of 10$^{-29}$ e~cm as a function of $\theta_u$ and $\theta_7$ in the case with $m_{H^0_2}=280$, $m_{H^\pm}=230$, $|\lam_7|=0.3$, $|\zeta_u|=0.01$, $|\zeta_d|=0.1$ and $|\zeta_e|=0.5$.
				The left (right) panel corresponds to the case with $\theta_d=0$ and $\theta_e=0~(\pi/4)$.
				The green solid lines denote $d_e=0$ and the area between the green dashed lines is the allowed region against the constraint from the electron EDM.
				In the right panel, the point marked by the star $(\theta_u, \theta_7)=(1.2,-1.8)$ is the benchmark point.
			}
			\label{fg:thetathetacontour}
		\end{figure}

In Fig.~\ref{fg:thetathetacontour}, the electron EDM is shown by the contour plot as a function of $\theta_u$ and $\theta_7$ in the case with $m_{H^0_2}=280$, $m_{H^\pm}=230$, $|\lam_7|=0.3$, $|\zeta_u|=0.01$, $|\zeta_d|=0.1$ and $|\zeta_e|=0.5$.
The left (right) panel corresponds to the case with $\theta_d=0$ and $\theta_e=0~(\pi/4)$.
Similar to Fig.~\ref{fg:massmasscontour}, the green solid line denotes $d_e=0$, and the area within dashed lines is allowed by the constraint from the electron EDM.
It is seen that $|d_e|$ becomes maximal at around $(\theta_u, \theta_7)=(\pm\pi/2, \pm\pi/2)$ in the left panel.
We find that there are regions satisfying $d_e\simeq0$ away from the origin, in which the destructive interference works successfully.
In the right panel, the shape of the contour is shifted to the upper-right region from that of the left panel.
Similarly, we find the regions satisfying $d_e\simeq0$ away from the origin.
	%%%   fig: eEDM theta-theta contour

From the discussion above, it is clarified that CP-violating phases of $\mathcal{O}(1)$ are allowed under the constraints from $d_e$ and $d_n$, because of the destructive interference.
It is also shown that the masses of $\mathcal{O}(100)$ GeV of the additional Higgs bosons are required to realize the destructive effects.
We note that in the parameter region allowed by $d_e$ the neutron EDM is typically two orders of the magnitude smaller than the current upper limit.
We set the benchmark point given in Tab.~\ref{tb:THDMinput}.
	%%%   tab: inputs of THDM para
		\begin{table}
			\centering
			\begin{tabular}{cccccc}
				\hline
					$M=240$,
				&	$m_{H^0_2}=280$,
				&	$m_{H^0_3}=230$,
				&	$m_{H^\pm}=230$
				&	(in GeV).
				\\ \hline
					$|\zeta_u|=0.01$,
				&	$|\zeta_d|=0.1$,
				&	$|\zeta_e|=0.5$,
				&	$|\lam_7|=0.3$,
				&	$\lam_2=0.5$.
				\\ \hline
					$\theta_u=1.2$,
				&	$\theta_d=0$,
				&	$\theta_e=\pi/4$,
				&	$\theta_7=-1.8$
				&	(in rad).
				\\ \hline
			\end{tabular}
			\caption{Input values for the THDM at the scale $m_Z$.}
			\label{tb:THDMinput}
		\end{table}
This point is marked as the star in the right panel of Fig.~\ref{fg:thetathetacontour}.

%%%%%%%%%%     collider signals     %%%%%%%%%%
\section{Collider signals}\label{sc:collider}

%%%%%     overview     %%%%%
\subsection{Overview}
%%%%%%%%%%
As we have discussed in the above section, the masses of the additional Higgs bosons $H_2$, $H_3$ and $H^\pm$ have to be ${\cal O}(100)$ GeV in order to 
realize the cancellation of the contribution from the Barr-Zee diagrams.
On the other hand, the couplings of the SM-like Higgs boson $h(=H_1)$ are the same as those of the SM Higgs boson at tree level, because 
we have imposed the alignment limit to the Higgs potential; i.e., ${\cal R}= 1$.
Therefore, the phenomenology for the additional Higgs bosons can be important in order to test our scenario.
In what follows, we first discuss the production mechanism of the additional Higgs bosons at hadron colliders.
We then investigate their decays, in which we can extract the effect of CP-violating phases by looking at the decay products of these Higgs bosons.

\subsection{Productions of the additional Higgs bosons}

\begin{figure}[t]
\begin{center}
 \includegraphics[width=70mm]{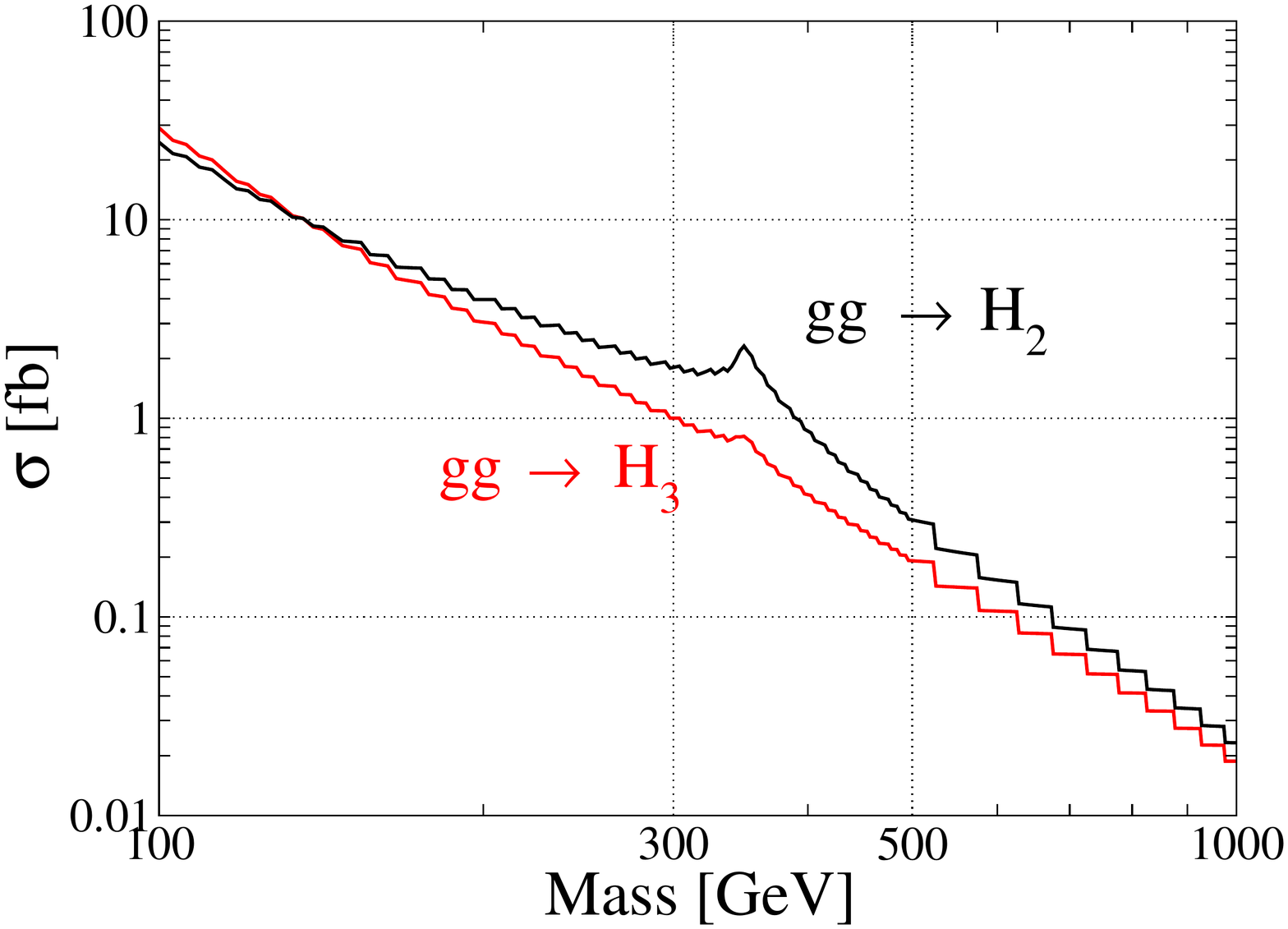}\hspace{5mm}
 \includegraphics[width=70mm]{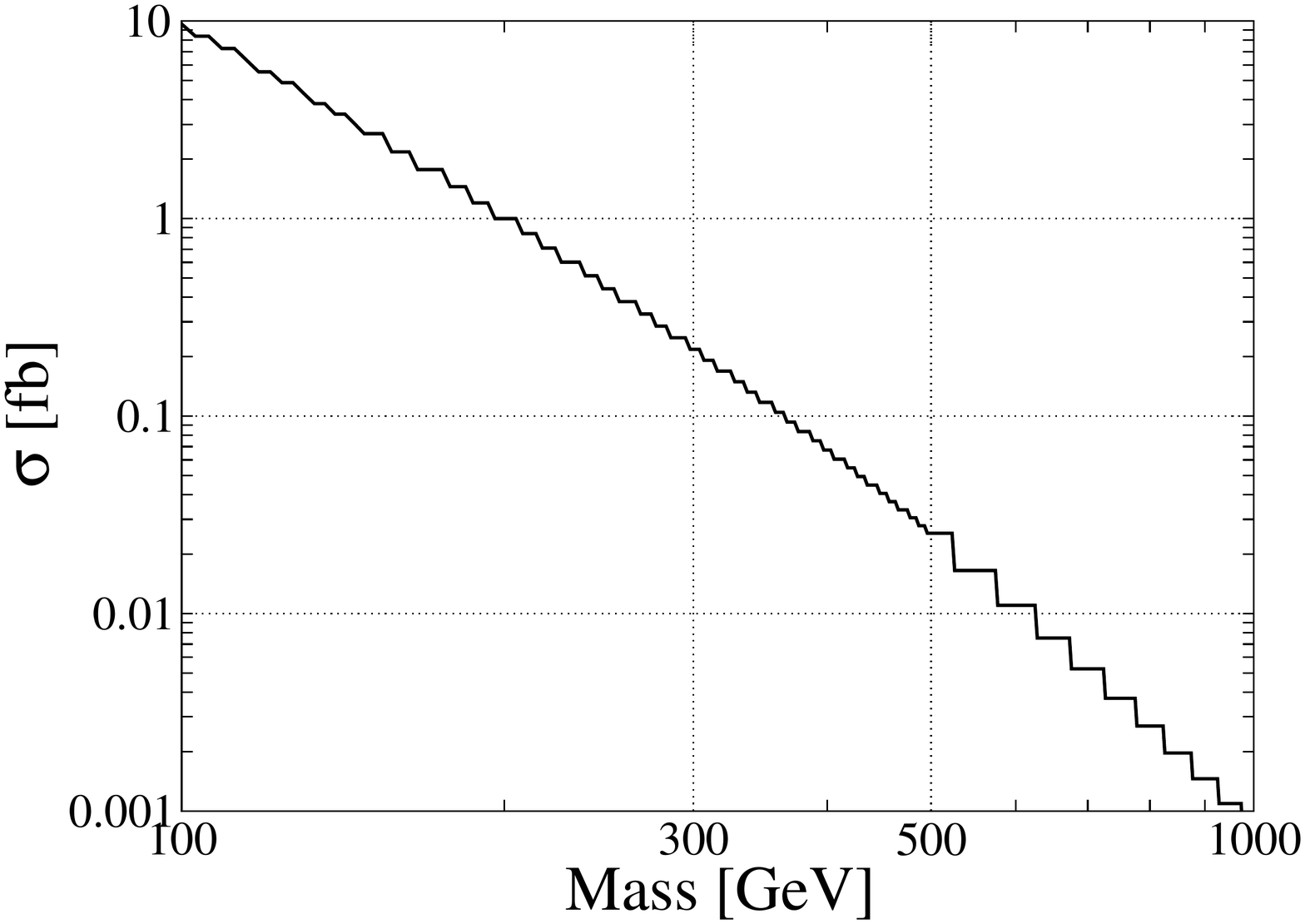}
   \caption{Cross section for $gg \to \phi^0$ (left) and $gg \to b\bar{b} \phi^0$ (right) at $\sqrt{s} = 13$ TeV ($\phi^0=H_2$ or $H_3$).
We take $|\zeta_u| = 0.01$, $|\zeta_d| = 0.1$, $\theta_u = 1.2$ and $\theta_d = 0$. }
   \label{ggf}
\end{center}
\vspace{7mm}
\end{figure}

In the scenario with the potential alignment, the $H_2VV$ and $H_3VV$ ($V=W,Z$) vertices vanish at tree level.
In addition, for the charged Higgs bosons the $H^\pm W^\mp Z$ vertex generally does not appear at tree level in THDMs~\cite{Grifols:1980uq,CapdequiPeyranere:1990qk,Kanemura:1997ej}, because we can define 
the Higgs basis, see Eqs.~\eqref{eq:basistransformation} and \eqref{eq:parametrizeddoublets}.
Furthermore, in general the $H^\pm W^\mp \gamma$ vertex does not appear at tree level in any kind of extended Higgs models, because of the $U(1)_{\text{em}}$ invariance.
The additional Higgs bosons can then be produced via the Yukawa interactions as follows: 
\begin{align}
&gg \to \phi^0 \quad \text{(gluon fusion)}, \label{xs1}\\
&gg \to t\bar{t}\phi^0 \quad \text{(top quark associated production )}, \label{xs2}\\
&gg \to b\bar{b}\phi^0 \quad \text{(bottom quark associated production )}, \label{xs3}\\
&gb \to H^\pm t \quad \text{(gluon-bottom fusion )}, \label{xs4}
\end{align}
where $\phi^0$ is $H_2$ or $H_3$.
They can also be produced in pair via the $s$-channel gauge boson mediations;
\begin{align}
&q\bar{q} \to Z^* \to H_2H_3, \label{xs5}\\
&q\bar{q}' \to W^* \to H^\pm \phi^0,\label{xs6} \\
&q\bar{q} \to Z^*/\gamma^* \to H^+H^-. \label{xs7}
\end{align}

Let us first consider the processes induced by the Yukawa interaction.
The dominant cross section for $\phi^0$ is provided from the gluon fusion process given in Eq.~(\ref{xs1}) whose value is estimated by 
\begin{align}
\sigma(gg \to \phi^0) = \sigma(gg \to h)_{\rm SM}\times \frac{\Gamma(\phi^0 \to gg)}{\Gamma(h \to gg)_{\rm SM}}, 
\end{align}
where $\sigma(gg \to h)_{\rm SM}$ is the cross section for  the SM Higgs boson, and $\Gamma(\phi^0 \to gg)$ [$\Gamma(h \to gg)_{\rm SM}$] is the 
decay rate of the $\phi^0 \to gg$ in our scenario [$h \to gg$ in the SM].
The value of $\sigma(gg \to h)_{\rm SM}$ can be referred from the LHC Higgs Cross Section Working Group~\cite{XSWG} at N3LO in QCD.
The cross sections for the top and bottom quark associated production can also be estimated by $\sigma_{t\bar{t}h}^{\rm SM}\times |\zeta_u|^2$ and 
$\sigma_{b\bar{b}h}^{\rm SM}\times |\zeta_d|^2$ with $\sigma_{t\bar{t}h}^{\rm SM}$ and $\sigma_{b\bar{b}h}^{\rm SM}$ being the cross section for $gg \to t\bar{t}h$  $gg \to b\bar{b}h$ in the SM, respectively.
According to \cite{XSWG}, $\sigma_{t\bar{t}h}^{\rm SM}$ and $\sigma_{b\bar{b}h}^{\rm SM}$ are almost the same value for a fixed value of $m_h$; e.g., 
they are given to be about 120 (100) fb at NLO in QCD, where $m_h= 200$ GeV and $\sqrt{s} = 13$ TeV are taken.
Thus, the cross section for the top quark associated production is negligibly small, because we have taken $|\zeta_u| = 0.01$ as the benchmark.
The cross sections for the gluon fusion and the bottom quark associated processes are shown in Fig.~\ref{ggf}.
It is seen that the cross section of the gluon fusion process is typically one order of the magnitude larger than that of the bottom quark associated process.

\begin{figure}[t]
\begin{center}
 \includegraphics[width=80mm]{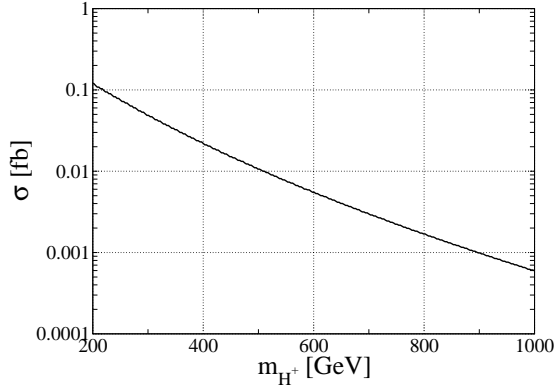}
   \caption{Cross section for $gb \to H^-t$  at $\sqrt{s} = 13$ TeV.
We take $|\zeta_u| = 0.01$, $|\zeta_d| = 0.1$, $\theta_u = 1.2$ and $\theta_d = 0$.}
   \label{gbh}
\end{center}
\vspace{7mm}
\end{figure}

For the charged Higgs bosons, if they are lighter than the top quark mass, they can be produced via the top decay $t\to H^\pm b$.
However, such light charged Higgs bosons have already been highly constrained by the current LHC data~\cite{CMS:2016szv,Aaboud:2018gjj}.
For instance, the upper limit on ${\cal B}(t \to H^\pm b)$ is given to be of order $10^{-3}$ at 95\% CL assuming  ${\cal B}(H^\pm \to \tau^\pm \nu) = 1$~\cite{Aaboud:2018gjj}.
We thus consider the case with the heavier charged Higgs bosons $m_{H^\pm} > m_t$, in which they can be produced via the gluon-bottom fusion process given in Eq.~(\ref{xs4}).
We evaluate the production cross section by using {\tt CalcHEP\_3.4.2}~\cite{Belyaev:2012qa} with the parton distribution functions of {\tt CTEQ6L}~\cite{Nadolsky:2008zw}.
In Fig.~\ref{gbh}, we show the production cross section of $gb \to H^\pm t$ process as a function of $m_{H^\pm}^{}$ at LO in QCD with $\sqrt{s} = 13$ TeV.
NLO corrections in QCD have been discussed in Refs.~\cite{Plehn:2002vy,Berger:2003sm}.

\begin{figure}[t]
\begin{center}
 \includegraphics[width=80mm]{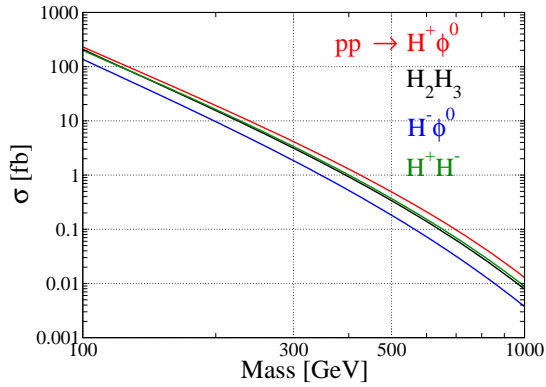}
   \caption{Cross sections for the pair production processes at $\sqrt{s} = 13$ TeV with $m_{H^\pm} = m_{H_2} = m_{H_3}$. }
   \label{pair}
\end{center}
\vspace{7mm}
\end{figure}

As mentioned above, there are pair production processes given in Eqs.~(\ref{xs5})--(\ref{xs7}).
Differently from the Yukawa induced processes, their cross section is determined by the gauge coupling for a fixed value of the mass of the additional Higgs boson.
Therefore, even when we take very small values of the $\zeta_f$ parameters, these production processes can be important.
Again, we use {\tt CalcHEP\_3.4.2} and {\tt CTEQ6L} for the evaluation of the cross section.
The cross section is shown in Fig.~\ref{pair}.
We see that these cross sections can be larger than those given by the Yukawa induced processes in the wide range of the mass region.
%%%%%%%%%%

%%%%%     decays     %%%%%
\subsection{Decays of the additional Higgs bosons}
We discuss the branching ratios of the additional Higgs bosons.
In the alignment limit $\mathcal{R}_{jk}=\delta_{jk}$, the additional Higgs bosons can mainly decay into a fermion pair.
They can also decay into a (off-shell) gauge boson and another Higgs boson if it is kinematically allowed.
In addition to these decay modes, we can consider loop-induced decay processes such as $H^0_{2,3}\to \gamma\gamma, Z\gamma, gg$ and $H^\pm\to W^\pm Z, W^\pm\gamma$.
Except for the $H^0_{2,3}\to gg$, the branching ratios of loop-induced processes are negligibly small; i.e., $BR(H^0_{2,3}\to \gamma\gamma/Z\gamma)$ and $BR(H^\pm\to W^\pm Z, W^\pm\gamma)$\cite{Kanemura:1997ej} are typically smaller than $\mathcal{O}(10^{-4})$.
In our benchmark point given in Tab.~\ref{tb:THDMinput}, the main decay modes of the additional Higgs bosons are given as follows
	%%%   main decay modes
		\eq{
			H^0_2	&\to b\bar{b}, c\bar{c}, \tau^+\tau^-, gg, W^\pm{}^* H^\mp, Z^*H^0_3
			\label{eq:decaymodes}
		,\\	H^0_3	&\to b\bar{b}, c\bar{c}, \tau^+\tau^-, gg
			\label{eq:decaymodes3}
		,\\	H^\pm	&\to tb, \tau^\pm\nu
			\label{eq:decaymodespm}
		.}
The explicit analytic formulae for the above decay rates are given in Appendix~\ref{sc:decayrate}.

In Fig.~\ref{fg:BR}, we show the branching ratios of $H^0_2$ (left), $H^0_3$ (center) and $H^\pm$ (right) as a function of $|\zeta_e|$ with the fixed parameters given in Tab.~\ref{tb:THDMinput}.
In this case, the upper limit of $|\zeta_e|$ is given to be $0.58$ from the constraint of $d_e$, which is denoted by the vertical solid line.
For the decay of $H^0_2$, $H^0_2\to W^\pm{}^*H^\mp$ and $H^0_2\to Z^*H^0_3$ are dominant in the wide range of $|\zeta_e|$.
On the other hand, $H^0_2\to\tau\tau$ can be important in the case with larger values of $|\zeta_e|$, whose branching ratio can be maximally about $20 \%$.
For the decay of $H^0_3$, the behavior of $BR(H^0_3\to\tau\tau)$ is similar to that of $H^0_2$.
However, because of the absent of the decays into a gauge boson and a Higgs boson the value of $BR(H^0_3\to\tau\tau)$ increases as compared with $BR(H^0_2\to\tau\tau)$.
In fact, it can be almost $100 \%$ with $|\zeta_e|\gtrsim0.5$.
For the decay of $H^\pm$, it is seen that $BR(H^\pm\to \tau^\pm\nu)$ can be larger than $BR(H^\pm\to tb)$ with $|\zeta_e|\gtrsim 0.4$.
	%%%   fig: branching ratio
		\begin{figure}
			\begin{tabular}{c}
				%%%
				\begin{minipage}{0.33\hsize}
					\centering
					\includegraphics[width=50 mm]{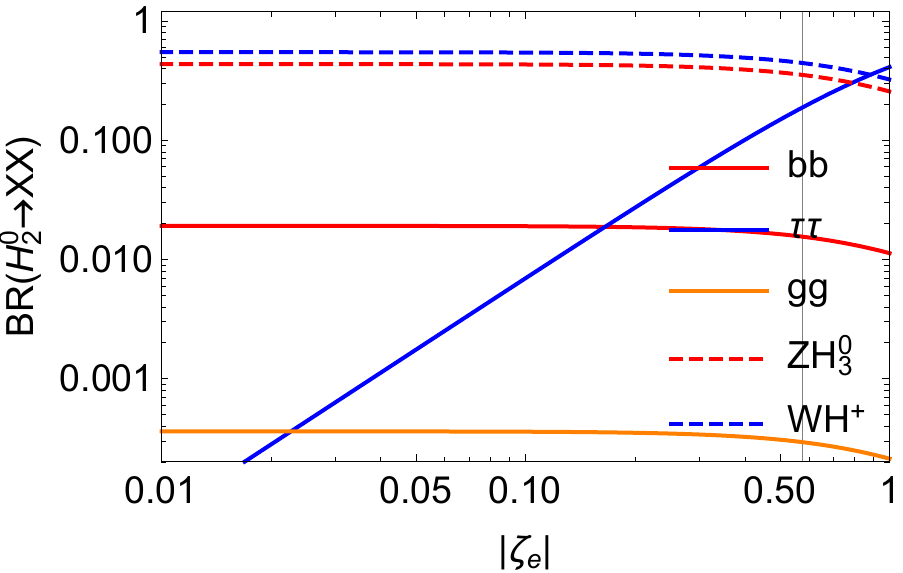}
					%\\(\ref{fg:BR}-1)
				\end{minipage}
				%%%
				\begin{minipage}{0.33\hsize}
					\centering
					\includegraphics[width=50 mm]{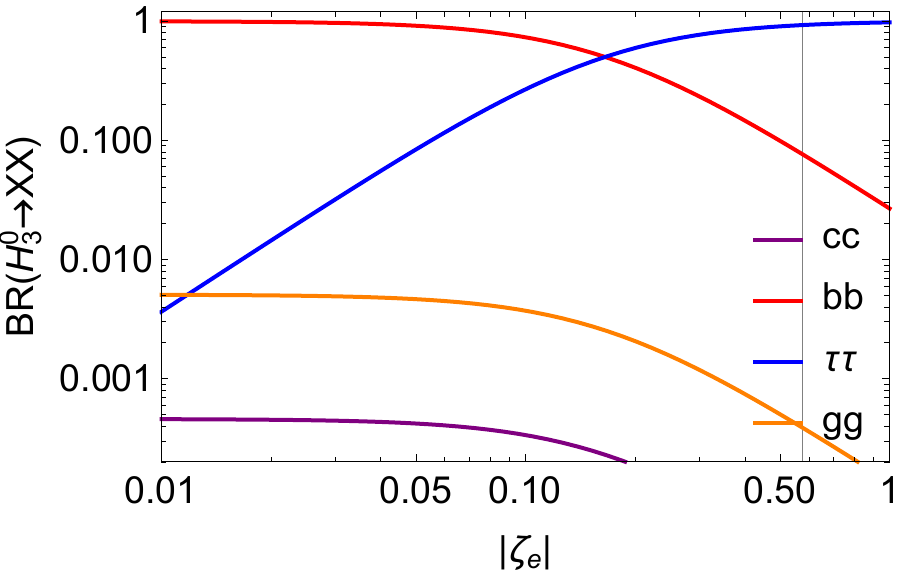}
					%\\(\ref{fg:BR}-2)
				\end{minipage}
				%%%
				\begin{minipage}{0.33\hsize}
					\centering
					\includegraphics[width=50 mm]{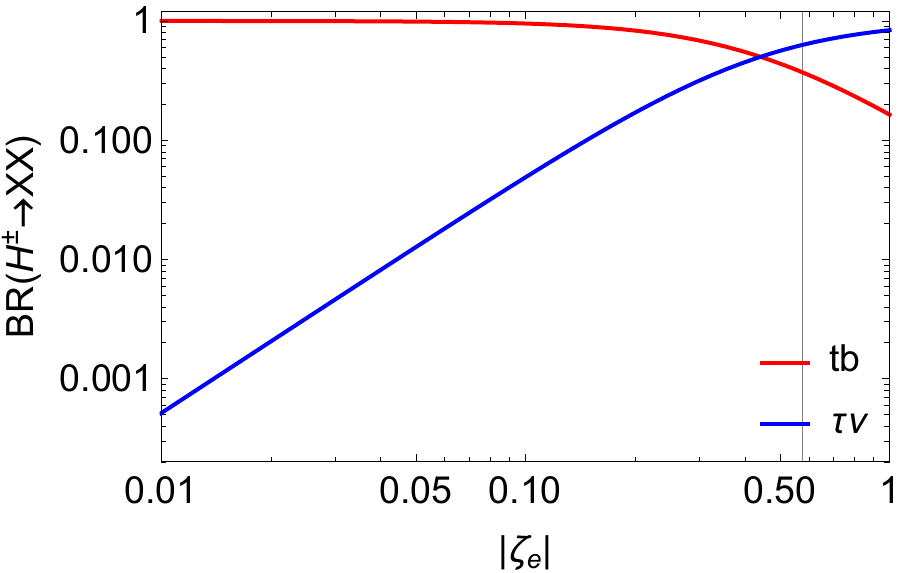}
					%\\(\ref{fg:BR}-3)
				\end{minipage}
				%%%
			\end{tabular}
			\caption{	
				Branching ratios of $H^0_2$ (left), $H^0_3$ (center) and $H^\pm$ (right) as the function of $|\zeta_e|$ with the fixed parameters given in Tab.~\ref{tb:THDMinput}.
				The vertical line denotes the upper limit $|\zeta_e|>0.58$ by the electron EDM.
			}
			\label{fg:BR}
		\end{figure}
Therefore, $H^0_{2,3}\to\tau\tau$ and $H^\pm\to\tau^\pm\nu$ can be important in our scenario.

%%%%%     angular distribution     %%%%%
\subsection{Angular distribution}
If the additional Higgs bosons are produced, the CP-violating phases can be measured from their decay products at collider experiments.
In our scenario, the CP-violating phases from the Yukawa interaction affect angular distributions of such decay products.
On the other hand, the effect of the CP-violating phases from the Higgs potential do not directly appear in the decays of the extra Higgs bosons into fermions.
However, if we can measure the CP-violating effect in the angular distributions, it indirectly proves the existence of the CP-violating phases in the Higgs potential because of the necessity of the cancellation of the EDM as discussed in Sec.~\ref{sc:EDM}.

We discuss the decay of the extra neutral Higgs bosons into a pair of $\tau$, because it provides relatively clearer signatures than the others and it can be dominant in our scenario, see Fig.~\ref{fg:BR}.
As discussed in Refs.~\cite{Kuhn:1982di,Hagiwara:2012vz,Harnik:2013aja,Jeans:2018anq}, the hadronic decays of $\tau$ can be useful to extract the CP-violating effects due to their simple kinematic structure.
We thus consider $H^0_j\to\tau^{-}\tau^{+}\to h^{-}\nu h^{+}\bar{\nu}$, where $h^{\pm}$ are hadrons, for instance, $\rho^{\pm}$ or $\pi^{\pm}$ mesons.

\begin{figure}
			\begin{tabular}{c}
				%%%
				\begin{minipage}{0.33\hsize}
					\centering
					\includegraphics[height=55 mm]{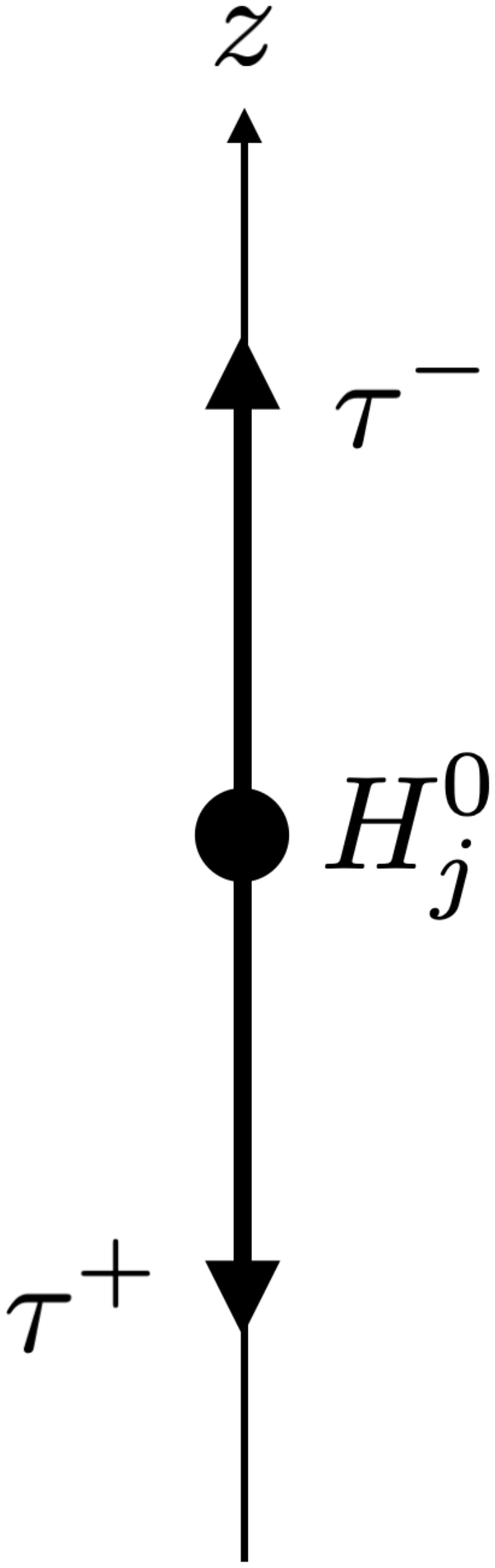}
					\\(\ref{fg:angleDef}-1) $H^0_j$ rest frame
				\end{minipage}
				%%%
				\begin{minipage}{0.33\hsize}
					\centering
					\includegraphics[height=55 mm]{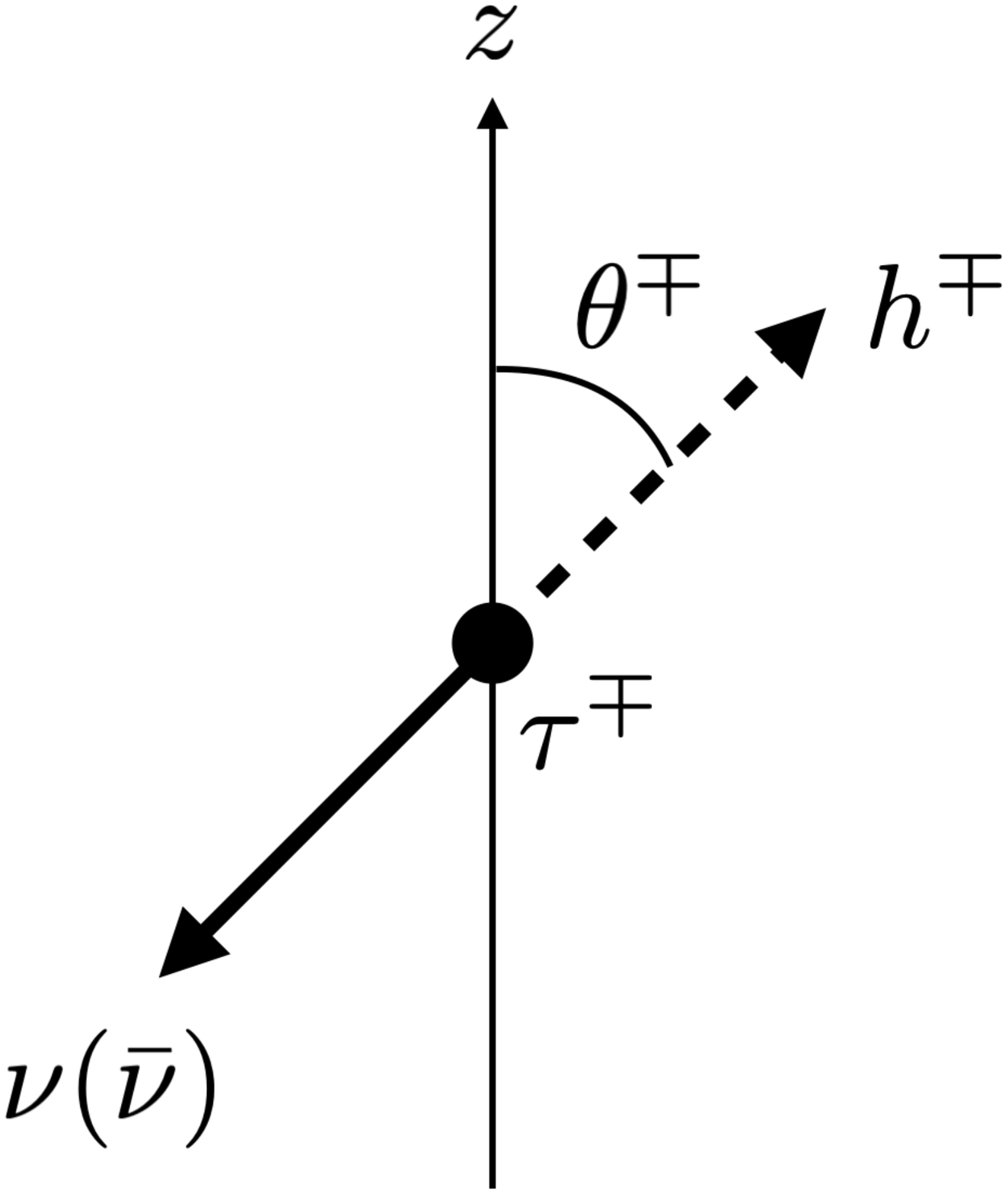}
					\\(\ref{fg:angleDef}-2) $\tau^\mp$ rest frame
				\end{minipage}
				%%%
				\begin{minipage}{0.33\hsize}
					\centering
					\includegraphics[height=55 mm]{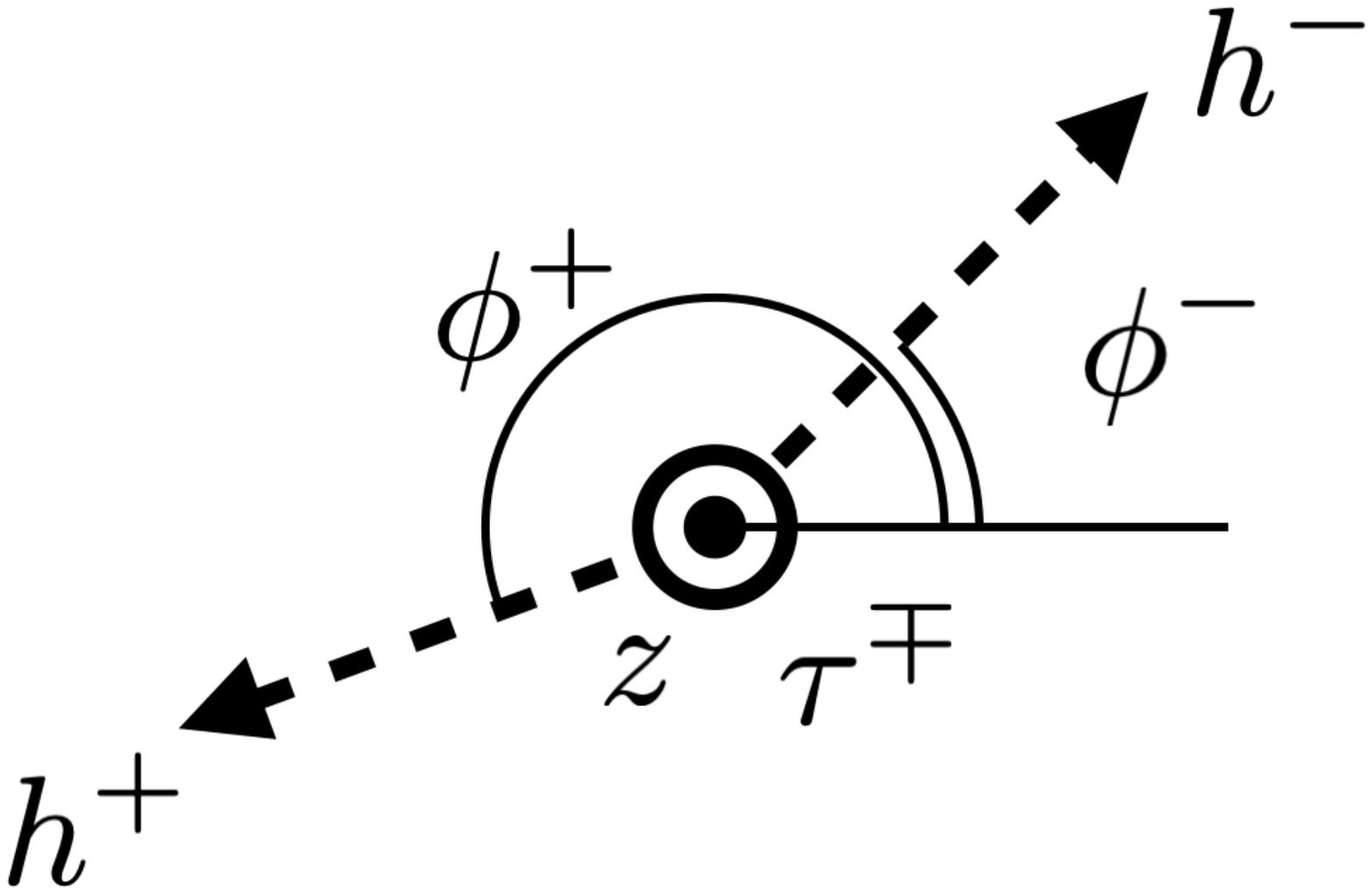}
					\\(\ref{fg:angleDef}-3) view looking along the $z$ axis
				\end{minipage}
				%%%
			\end{tabular}
			\caption{
				Schematic pictures for the angles $\theta^\pm$ and $\phi^\pm$ defined on the rest frame of $\tau^\pm$.
				The $z$ axis is defined along with the direction of $\tau^-$ on the rest frame of $H^0_j$.
				}
			\label{fg:angleDef}
		\end{figure}

\begin{figure}
			\begin{tabular}{c}
				\begin{minipage}{0.5\hsize}
					\centering
					\includegraphics[width=75 mm]{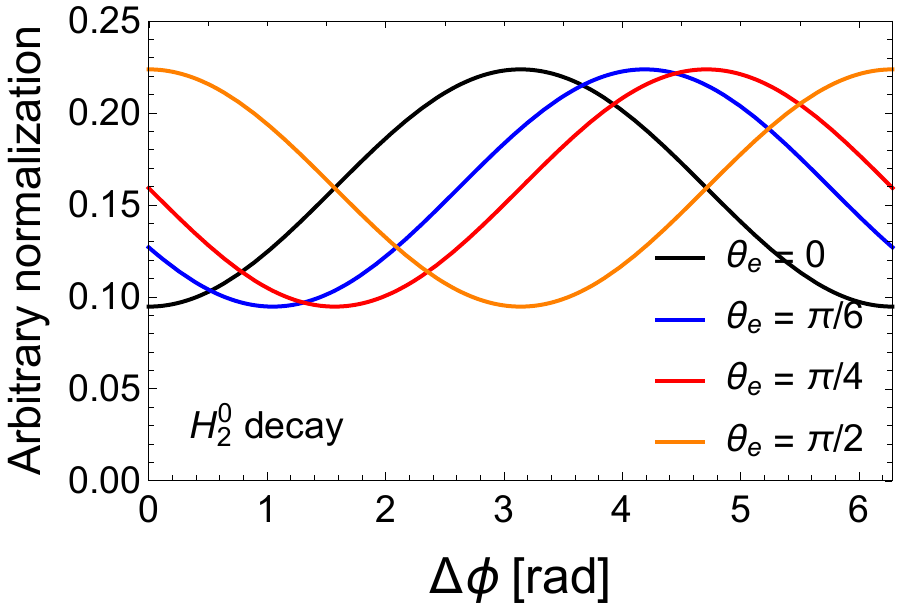}
					%\\(\ref{fg:angldistplot}-1)
				\end{minipage}
				%%%
				\begin{minipage}{0.5\hsize}
					\centering
					\includegraphics[width=75 mm]{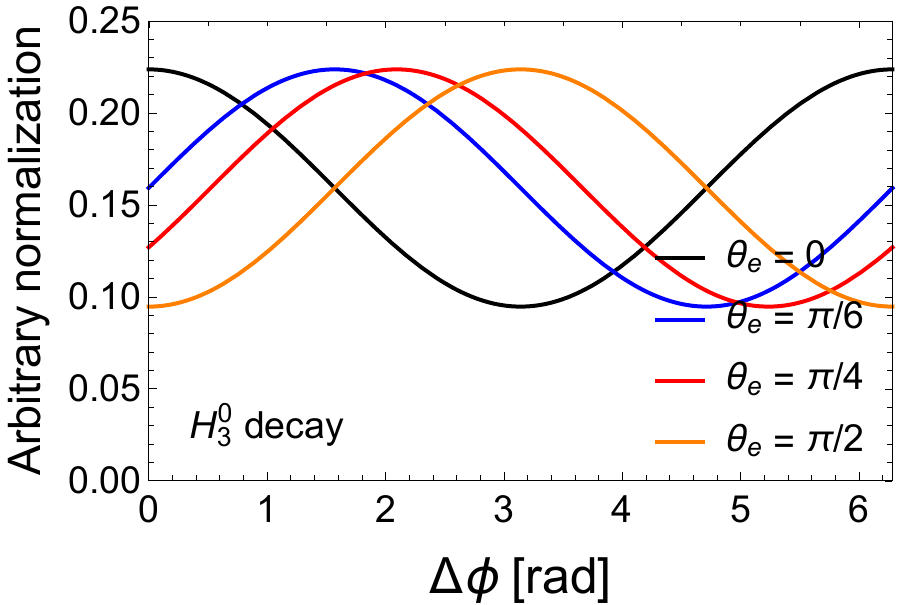}
					%\\(\ref{fg:angldistplot}-2)
				\end{minipage}
				%%%
			\end{tabular}
			\caption{Normalized $\Delta\phi$ distributions for $H^0_j\to\tau^{-}\tau^{+}\to h^{-}\nu h^{+}\bar{\nu}$ with $j=$2 (left) and 3 (right).}
			\label{fg:angldistplot}
		\end{figure}

The average of the squared amplitude for $H^0_j\to\tau^{-}\tau^{+}\to h^{-}\nu h^{+}\bar{\nu}$ is calculated as
	%%%   amplitude
		\begin{align}
			\overline{|\mathcal{M}(\theta^-, \phi^-, \theta^+, \phi^+)|^{2}}\propto(1+\cos\theta^-\cos\theta^+)-\sin\theta^-\sin\theta^+\cos(2\arg[\kappa_e^j]+\phi^- -\phi^+)
		\label{eq:amplitude}
		,\end{align}
where the mass of $\tau$ is neglected, and the angles $\theta^\pm$ and $\phi^\pm$ are defined as the momentum direction of the meson on the rest frame of $\tau^\pm$ as depicted in Fig.~\ref{fg:angleDef}.
	%%%   fig: angle
By applying Eq.~\eqref{eq:amplitude} to the decays of $H^0_{2,3}$, the angular distribution of the decay products of $H^0_{2,3}$ are obtained as follows
	%%%   angular distribution
		\begin{align}
			&\int\int d\theta^- d\theta^+ ~\overline{|\mathcal{M}(H^0_2\to\tau^{-}\tau^{+}\to h^{-}\nu h^{+}\bar{\nu})|^{2}}
								\propto\pi^2-4\cos(2\theta_e-\Delta\phi)
		,\\	&\int\int d\theta^- d\theta^+ ~\overline{|\mathcal{M}(H^0_3\to\tau^{-}\tau^{+}\to h^{-}\nu h^{+}\bar{\nu})|^{2}}
								\propto\pi^2-4\cos(2\theta_e-\Delta\phi+\pi/2)
		,\end{align}
where $\Delta\phi\equiv\phi^+-\phi^-$.
From these expressions, we can extract $\theta_e$ by looking at the $\Delta\phi$ distributions.
In Fig.~\ref{fg:angldistplot}, we show the $\Delta\phi$ distributions in the decay of $H^0_2$ (left) and that of $H^0_3$ (right) for each fixed value of $\theta_e$.
Depending on the value of $\theta_e$, the shape of the distributions changes, so that we may be able to extract the information of $\theta_e$.
It goes without saying that dedicated studies with signal and background simulations have to be performed in order to know the feasibility of extracting the CP-violating phases in our scenario.
	%%%   fig: angldistplot

%%%%%%%%%%     RGE     %%%%%%%%%%
\section{Renormalization group equation analysis}\label{sc:RGE}
One might think that the destructive cancellation of the contributions from BZ diagrams seems to be a kind of artificial fine tuning to satisfy the data.
In order to see the stability of our scenario, we investigate the high energy behavior by the RGE analysis.
We discuss the scale dependence of the electron EDM $d_e$ in the case where the destructive interference sufficiently realizes at the scale of $m_Z$ so that the current data are satisfied.
We evaluate the running of all the dimensionless couplings from the scale of $m_Z$ to a high energy scale by using the one-loop $\beta$-functions given in Appendix~\ref{sc:betafunc}.

\begin{figure}
			\begin{tabular}{c}
				%%%
				\begin{minipage}{0.5\hsize}
					\centering
					\includegraphics[width=70 mm]{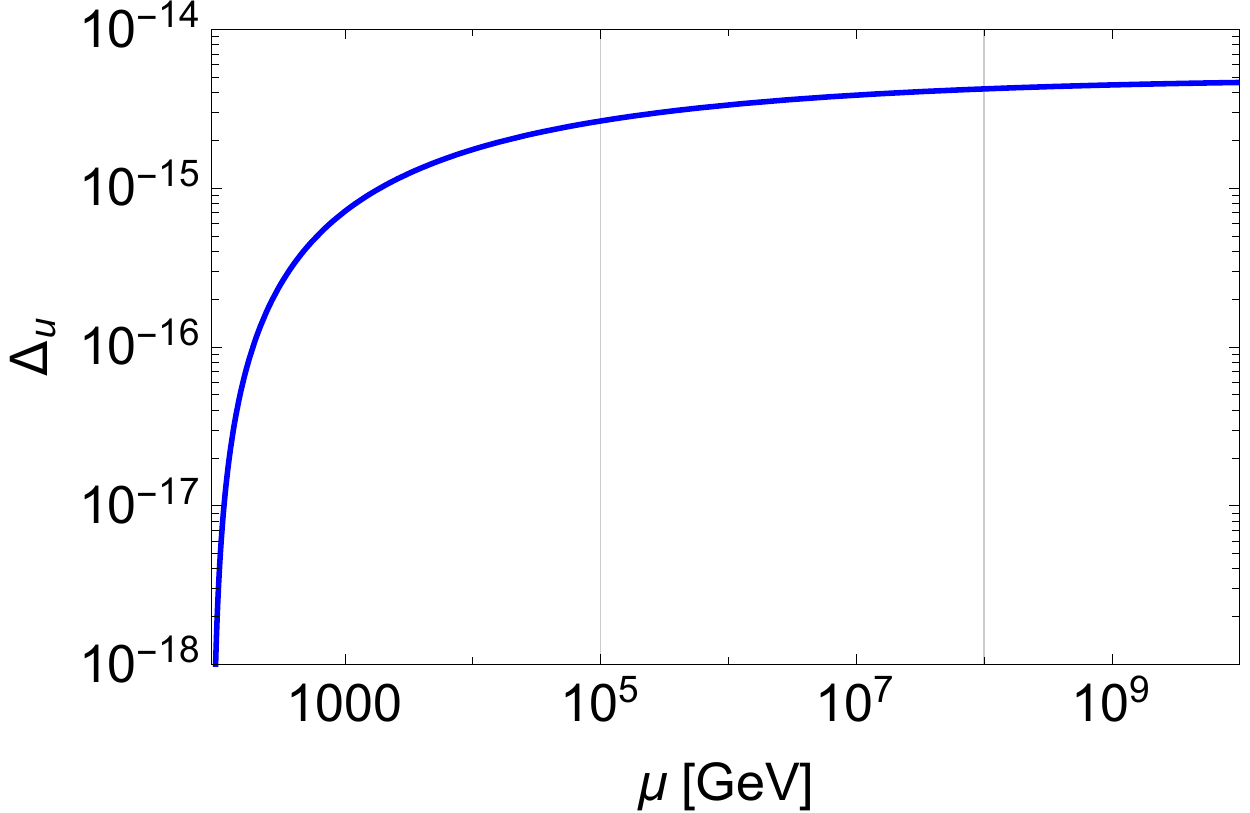}
					%\\(\ref{fg:Deltarun}-1)
				\end{minipage}
				%%%
				\begin{minipage}{0.5\hsize}
					\centering
					\includegraphics[width=70 mm]{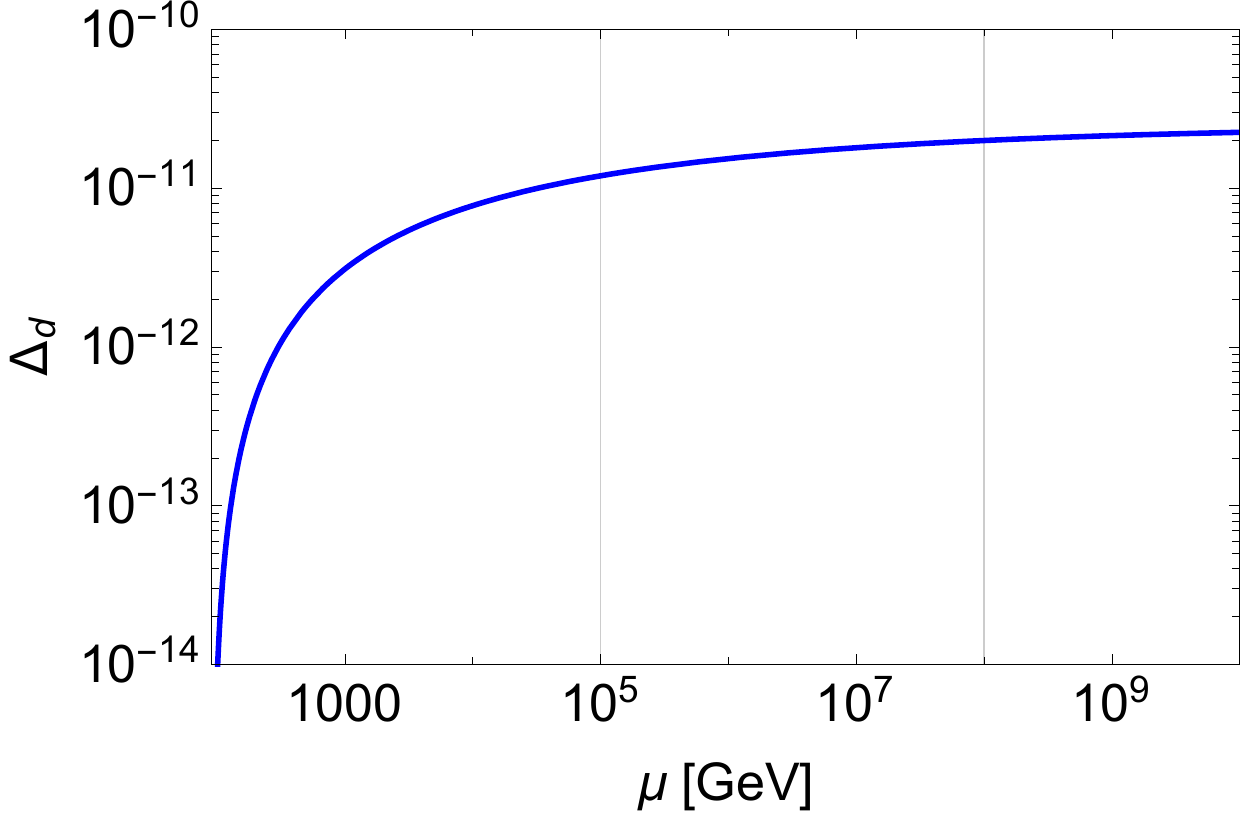}
					%\\(\ref{fg:Deltarun}-2)
				\end{minipage}
				%%%
			\end{tabular}
			\caption{RGE running of the parameter defined in Eq.~\eqref{eq:Delta} from $m_Z$ with the benchmark parameters given in Tab.~\ref{tb:THDMinput}}
			\label{fg:Deltarun}
		\end{figure}

In order to investigate the scale dependence of $d_e$, we first consider how the alignment of the Yukawa interaction can be broken at a high energy scale.
The magnitude of the departure from the alignment limit can be parametrized as\cite{Gori:2017qwg}
	%%%
		\begin{align}
			\Delta_q\equiv\Tr [\delta y_q^\dagger \delta y_q],	~~~\textrm{where}
		~~~	\delta y_q\equiv \hat{\rho}_q-\rbra{\frac{\hat{\rho}_q^{33}}{\hat{M}_q^{33}}}\hat{M}_q,	~~~(q=u,d)
		\label{eq:Delta}
		.\end{align}
We note that $\Delta_q$ vanishes at the alignment limit which is assumed at the scale of $m_Z$.
The behavior of the running of $\Delta_u$ (left) and $\Delta_d$ (right) is shown in Fig.~\ref{fg:Deltarun}.
It is clearly seen that both $\Delta_u$ and $\Delta_d$ are significantly small up to a high energy scale such as at least about $10^{10}$ GeV where the Landau pole appears (see the discussion below).
This is because the source of the breaking of the alignment mainly arises from the tiny off-diagonal elements of the CKM matrix.
Therefore, the Yukawa alignment approximately holds up to a high energy scale.
	%%%

\begin{figure}
			\begin{tabular}{c}
				%%%
				\begin{minipage}{0.5\hsize}
					\centering
					\includegraphics[width=70 mm]{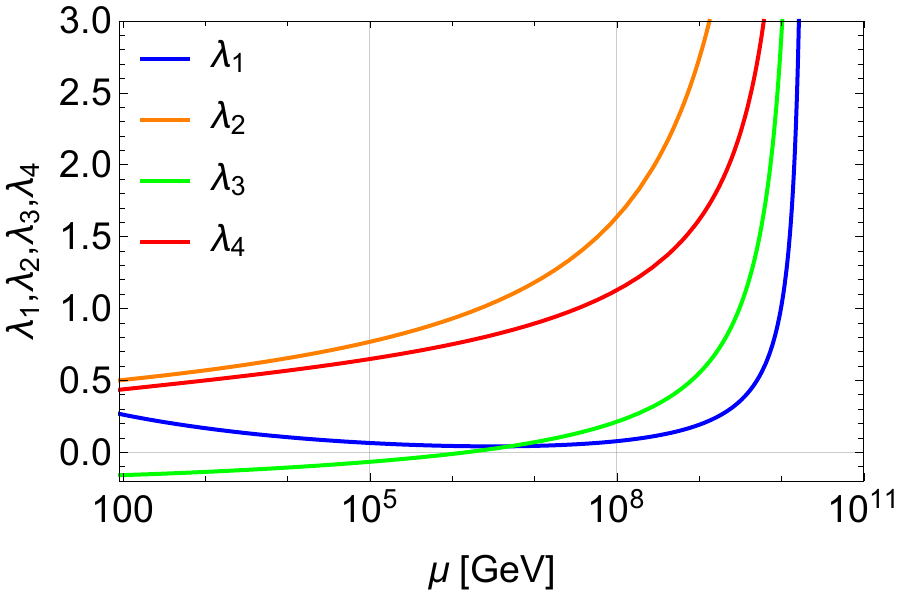}
					%\\(\ref{fg:lamrun}-1)
				\end{minipage}
				%%%
				\begin{minipage}{0.5\hsize}
					\centering
					\includegraphics[width=70 mm]{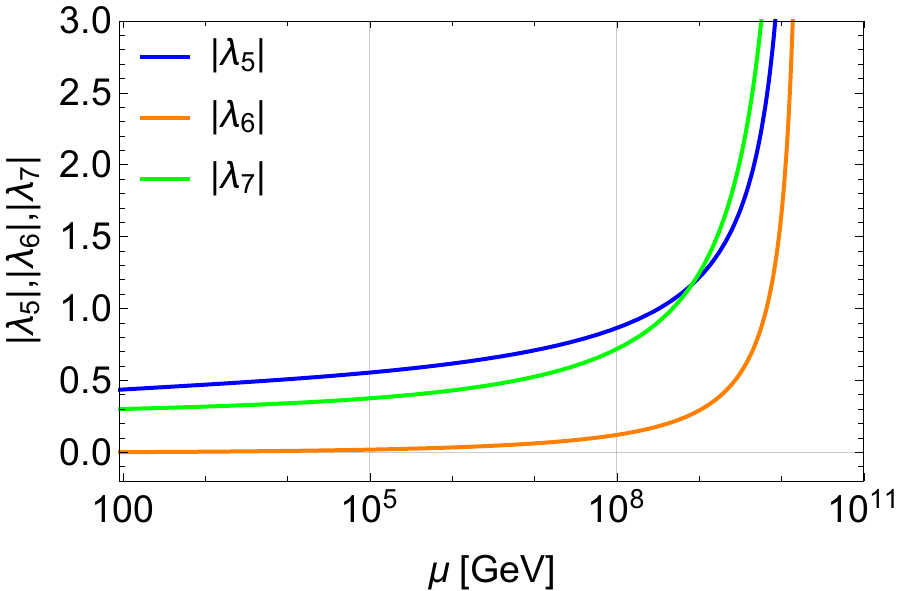}
					%\\(\ref{fg:lamrun}-2)
				\end{minipage}
				%%%
			\end{tabular}
			\caption{
				RGE-running behavior of the Higgs self couplings from $m_Z$ with the benchmark parameters given in Tab.~\ref{tb:THDMinput}.
				The left (right) panel corresponds to $\lam_{1\text{-}4}$ ($|\lam_{5\text{-}7}|$).
					}
			\label{fg:lamrun}
		\end{figure}

Next, we consider how the alignment of the Higgs potential can be broken at a high energy scale.
This can be clarified by looking at the running of the $\lam_6$ parameter, see Eq.~\eqref{eq:massmatrix}.
In Fig.~\ref{fg:lamrun}, we show the scale dependence of the magnitude of all the dimensionless parameters in the Higgs potential.
We can see that the $|\lam_6|$ parameter quite slowly increases as the scale is getting higher, and it blows up together with all the other couplings at 
around $\mu = 10^{10}$ GeV which is the scale appearing the Landau pole.
Therefore, our scenario is stable up to such a high energy scale.
	%%%

\begin{figure}
					\centering
					\includegraphics[width=70 mm]{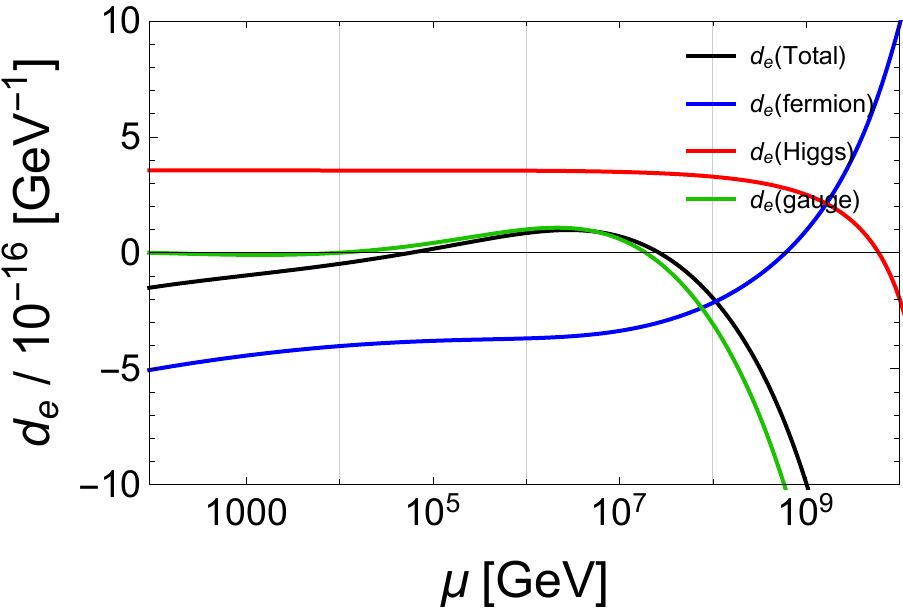}
			\caption{RGE running of the electron EDM from $m_Z$ with the benchmark parameters given in Tab.~\ref{tb:THDMinput}.}
			\label{fg:eEDMrun}
		\end{figure}

Finally, we discuss the scale dependence of $d_e$ which can be evaluated by
	%%%
		\begin{align}
			\mu\frac{\partial}{\partial\mu}\hat{d}_e(\mu)
				=	\mu\frac{\partial}{\partial\mu}\hat{d}_e(\textrm{fermion})
					+\mu\frac{\partial}{\partial\mu}\hat{d}_e(\textrm{Higgs})
					+\mu\frac{\partial}{\partial\mu}\hat{d}_e(\textrm{gauge})
		.\end{align}
We note that the contribution from the gauge boson loop $\hat{d}_e({\rm gauge} )$ appears at higher energy scales, because of the breaking of the potential alignment.
In addition, the SM-like Higgs boson $H_1^0$ is no longer the pure CP-even state at higher energy scales, so that $H_1^0$ can also but slightly contribute to $\hat{d}_e$ as well as $H_2^0$ and $H_3^0$.
In Fig.~\ref{fg:eEDMrun}, we show the scale dependence of each contribution to $\hat{d}_e$; i.e., $\hat{d}_e(\textrm{fermion})$, $\hat{d}_e(\textrm{Higgs})$ and $\hat{d}_e(\textrm{gauge})$, by taking into account the above mentioned issues, where we neglect subdominant contributions from non-BZ type diagrams.
We see that the cancellation between $\hat{d}_e({\rm fermion})$ and $\hat{d}_e({\rm Higgs})$ still works at higher energy scales.
However, because of the appearance of $\hat{d}_e({\rm gauge})$, the total value is getting larger, and it exceeds $|d_e|=1.747\times10^{-16}\textrm{~GeV}^{-1}$ at around $\mu = 10^8$ GeV\footnote{Although $\hat{d_e}$ at the high energy scale is not constrained by experiments, it would be viable to see the stability of the parameter set to realize the destructive cancellation at the $m_Z$ scale.}.
	%%%

%%%%%%%%%%     discussions and conclusions     %%%%%%%%%%
\section{Discussions and Conclusions}\label{sc:summary}

We have discussed the general THDM with the multiple sources of CP-violating phases in the Yukawa interaction and the Higgs potential.
In order to avoid the FCNCs at tree-level, we have imposed the Yukawa alignment.
In addition, the alignment in the Higgs potential is imposed in order that coupling constants of the Higgs boson with the mass of 125 GeV with SM particles 
are the same as those of the SM Higgs boson at tree level.

In this framework, we have computed the contributions from the BZ type diagrams to the electron and neutron EDMs.
We have found that there are non-trivial regions of the parameter space with ${\cal O}(1)$ CP-violating phases allowed by the current bounds from the EDM searches
if the masses of the additional Higgs bosons are taken to be ${\cal O}(100)$ GeV.
Such non-trivial solutions to satisfy the EDM bounds are obtained due to the destructive interference between the fermion and the Higgs boson loop contributions in the BZ diagram.

We then have considered how our scenario can be tested at collider experiments.
In particular, we have focused on the decays of the additional neutral Higgs bosons into a pair of tau leptons with their hadronic decays.
The difference of the azimuthal angles between the two hadron jets strongly depends on the CP-violating phases of the Yukawa interaction for the decaying Higgs boson.
Therefore, measurements of such a specific distribution can be the direct test of the CP-violating phases in the Yukawa interaction.
Furthermore, this can be the indirect test of the CP-violating phase in the Higgs potential, because of the necessity of the destructive cancellation of the EDM.
We have shown the angular distributions for several values of the CP-violating phases of the Yukawa interaction.

Finally, we have discussed the scale dependence of the dimensionless couplings by using the RGEs at one-loop level.
We have confirmed that both the alignment in the Yukawa interaction and that in the Higgs potential can be stable up to a high energy scale such as $10^{8}$ GeV.

Before closing this paper, we give a brief comment on the possibility of EWBG in our scenario.
At the zero temperature, the VEVs of the Higgs doublets are taken to be $(\eval{\Phi_1^0},\eval{\Phi_2^0})=(v/\sqrt{2},0)$, while 
at the finite temperature, they can be $(\eval{\Phi_1^0},\eval{\Phi_2^0})=(v_1',v_2')/\sqrt{2}$~\cite{Cline:2011mm}.
If this kind of the structure of the phase transition can be realized, the complex phase can appear in the top quark mass via $\zeta_u$ during the EWPT, which may be able to generate the baryon asymmetry of the Universe.
In this case, our scenario discussed in this paper is important for successful and tastable models for EWBG.

%%%%%%%%%%     acknowledgments     %%%%%%%%%%
\begin{acknowledgments}
The work of S. K. was supported in part by Grant-in-Aid for Scientific Research on Innovative Areas, the Ministry of Education, Culture, Sports, Science and Technology, No.~16H06492 and No.~18H04587 and also by JSPS, Grant-in-Aid for Scientific Research, Grant No.~18F18022, No.~18F18321 and No.~20H00160.
The work of K. Y. was supported in part by the Grant-in-Aid for Early-Career Scientists, No.~19K14714.
\end{acknowledgments}

%%%%%%%%%%     appendix     %%%%%%%%%%
\newpage
\appendix
\section{Parameters of the scalar potential: relations between two bases}\label{sc:parameters}
The relations between the potential parameters on the original basis and the Higgs basis given in Eqs.~\eqref{eq:potential1} and \eqref{eq:potential2} in Sec.~\ref{sc:model} are expressed as follows\cite{Davidson:2005cw}
	%%%   coefficients
		\begin{alignat}{2}
			\mu_1^2&=&&+\primed{\mu}_1^2c_\beta^2+\primed{\mu}_2^2s_\beta^2+2\re[\primed{\mu}_3^2 e^{i\xi}]c_\beta s_\beta,
		\\	\mu_2^2&=&&+\primed{\mu}_1^2s_\beta^2+\primed{\mu}_2^2c_\beta^2-2\re[\primed{\mu}_3^2 e^{i\xi}]c_\beta s_\beta,
		\\	\mu_3^2&=&&-(\primed{\mu}_1^2-\primed{\mu}_2^2)c_\beta s_\beta
							+\re[\primed{\mu}_3^2 e^{i\xi}](c_\beta^2-s_\beta^2)+i\im[\primed{\mu}_3^2 e^{i\xi}],
		\\	\lam_1&=&&+\primed{\lam}_1c_\beta^4+\primed{\lam}_2s_\beta^4
							+2\primed{\lam}_{345}c_\beta^2s_\beta^2
							+4\rbra{\re[\primed{\lam}_6 e^{i\xi}]c_\beta^2+\re[\primed{\lam}_7 e^{i\xi}]s_\beta^2}c_\beta s_\beta,
		\\	\lam_2&=&&+\primed{\lam}_1s_\beta^4+\primed{\lam}_2c_\beta^4
							+2\primed{\lam}_{345}c_\beta^2s_\beta^2
							-4\rbra{\re[\primed{\lam}_6 e^{i\xi}]s_\beta^2+\re[\primed{\lam}_7 e^{i\xi}]c_\beta^2}c_\beta s_\beta,
		\\	\lam_3&=&&+\primed{\lam}_3+(\primed{\lam}_1+\primed{\lam}_2-2\primed{\lam}_{345})c_\beta^2s_\beta^2
							-2\rbra{\re[\primed{\lam}_6 e^{i\xi}]-\re[\primed{\lam}_7 e^{i\xi}]}(c_\beta^2-s_\beta^2)c_\beta s_\beta,
		\\	\lam_4&=&&+\primed{\lam}_4+(\primed{\lam}_1+\primed{\lam}_2-2\primed{\lam}_{345})c_\beta^2s_\beta^2
							-2\rbra{\re[\primed{\lam}_6 e^{i\xi}]-\re[\primed{\lam}_7 e^{i\xi}]}(c_\beta^2-s_\beta^2)c_\beta s_\beta,
		\\	\lam_5&=&&+\re[\primed{\lam}_5 e^{2i\xi}]
							+(\primed{\lam}_1+\primed{\lam}_2-2\primed{\lam}_{345})c_\beta^2s_\beta^2
							-2\rbra{\re[\primed{\lam}_6 e^{i\xi}]-\re[\primed{\lam}_7 e^{i\xi}]}(c_\beta^2-s_\beta^2)c_\beta s_\beta
							\nonumber
							\\&&&+i\sbra{\im[\primed{\lam}_5 e^{2i\xi}](c_\beta^2-s_\beta^2)
							-2\rbra{\im[\primed{\lam}_6 e^{i\xi}]-\im[\primed{\lam}_7 e^{i\xi}]}c_\beta s_\beta}v^2,
		\\	\lam_6&=&&-\sbra{\primed{\lam}_1c_\beta^2-\primed{\lam}_2s_\beta^2
							-\primed{\lam}_{345}(c_\beta^2-s_\beta^2)}c_\beta s_\beta
							+\re[\primed{\lam}_6 e^{i\xi}]c_\beta^2(c_\beta^2-3s_\beta^2)
							-\re[\primed{\lam}_7 e^{i\xi}]s_\beta^2(s_\beta^2-3c_\beta^2)
							\nonumber
							\\&&&+i\rbra{\im[\primed{\lam}_5 e^{2i\xi}]c_\beta s_\beta+\im[\primed{\lam}_6 e^{i\xi}]c_\beta^2+\im[\primed{\lam}_7 e^{i\xi}]s_\beta^2},
		\\	\lam_7&=&&-\sbra{\primed{\lam}_1s_\beta^2-\primed{\lam}_2c_\beta^2
							+\primed{\lam}_{345}(c_\beta^2-s_\beta^2)}c_\beta s_\beta
							-\re[\primed{\lam}_6 e^{i\xi}]s_\beta^2(s_\beta^2-3c_\beta^2)
							+\re[\primed{\lam}_7 e^{i\xi}]c_\beta^2(c_\beta^2-3s_\beta^2)
							\nonumber
							\\&&&-i\rbra{\im[\primed{\lam}_5 e^{2i\xi}]c_\beta s_\beta-\im[\primed{\lam}_6 e^{i\xi}]s_\beta^2-\im[\primed{\lam}_7 e^{i\xi}]c_\beta^2}
		,\end{alignat}
where $\xi\equiv\xi_2-\xi_1$, $c_\beta=\cos\beta$, $s_\beta=\sin\beta$ and $\primed{\lam}_{345}\equiv\primed{\lam}_3+\primed{\lam}_4+\re[\primed{\lam}_5 e^{2i\xi}]$.

\section{Exact formulae of the Barr-Zee type contributions}\label{sc:BZformulae}
We here list the Barr-Zee type contributions to the EDM (CEDM) for a fermion (quark) $d_f^\gamma$, $d_f^Z$ and $d_f^W$ $(d_q^C)$ given in Eqs.~\eqref{eq:BZcontribution1} and \eqref{eq:BZcontribution2} (Eq.~\eqref{eq:nEDM}) in Sec.~\ref{sc:EDM}.
The fermion-loop contributions to the EDM are
	%%%   fermion-loop BZ
		\small
		\begin{align}
				d_f^V(f')	&=\frac{em_f}{(16\pi^2)^2}8g_{Vff}^vg_{Vf'f'}^v
											Q_{f'}N_C\frac{m_{f'}^2}{v^2}
									\nn\\&\quad\quad
									\times\sum_j^3\int_0^1dz\Bigg\{
												\im[\kappa_f^j]\re[\kappa_{f'}^j]
												\rbra{\frac{1}{z}-2(1-z)}
												+\re[\kappa_f^j]\im[\kappa_{f'}^j]
												\frac{1}{z}
											\Bigg\}
									C^{VH^0_j}_{f'f'}(z)
				,\\
				d_{f(I_f=-\frac{1}{2})}^W(tb)&=\frac{eg_2^2m_f}{(16\pi^2)^2}N_C%|V_{tb}|^2
									\int_0^1dz\Bigg\{
											\frac{m_t^2}{v^2}\im[\zeta_f^*\zeta_u]\frac{2-z}{z}
											+\frac{m_b^2}{v^2}\im[\zeta_f^*\zeta_d]
									\Bigg\}[Q_t(1-z)+Q_bz]
									C^{WH^\pm}_{tb}(z)
				,\\
				d_{f(I_f=+\frac{1}{2})}^W(tb)&=\frac{eg_2^2m_f}{(16\pi^2)^2}N_C%|V_{tb}|^2
									\int_0^1dz\Bigg\{
											\frac{m_t^2}{v^2}\im[\zeta_f\zeta_u^*]
											+\frac{m_b^2}{v^2}\im[\zeta_f\zeta_d^*]\frac{1+z}{1-z}
									\Bigg\}[Q_t(1-z)+Q_bz]
									C^{WH^\pm}_{tb}(z)
		,\end{align}
		\normalsize
the Higgs boson-loop contributions to EDM are
	%%%   scalar-loop BZ
		\begin{align}
				d_f^V(H^\pm)		&=\frac{em_f}{(16\pi^2)^2}4g_{Vff}^v(ig_{H^+H^-V})
									\sum_j^3\im[\kappa_f^j]\frac{g_{H^\pm H^\mp H^0_j}}{v}
										\int_0^1dz(1-z)
									C^{VH^0_j}_{H^{^\pm}H^{^\pm}}(z)
				,\\
				d_f^W(H^\pm H^0)		&=\frac{eg_2^2m_f}{(16\pi^2)^2}\frac{1}{2}
									\sum_j^3\im[\kappa_f^j]\frac{g_{H^{^\pm} H^{^\mp} H^0_j}}{v}
										\int_0^1dz(1-z)
										C^{WH^\pm}_{H^{^\pm}H^0_j}(z)
		,\end{align}
the gauge-loop contributions to EDM are
	%%%   gauge-loop BZ
		\small
		\begin{align}
				d_f^V(W)		&=\frac{em_f}{(16\pi^2)^2}8g_{Vff}^v(ig_{WWV})\frac{m_W^2}{v^2}
									\sum_j^3\mathcal{R}_{1j}\im[\kappa_f^j]
									\nn\\&\quad\quad
										\times\int_0^1dz\Bigg[
											\cbra{
												\rbra{6-\frac{m_V^2}{m_W^2}}
												+\rbra{1-\frac{m_V^2}{2m_W^2}}\frac{m_{H^0_j}^2}{m_W^2}
											}\frac{(1-z)}{2}
											-\rbra{4-\frac{m_V^2}{m_W^2}}\frac{1}{z}
										\Bigg]
									C^{VH^0_j}_{WW}(z)
				,\\
				d_f^W(WH^0)	&=\frac{eg_2^2m_f}{(16\pi^2)^2}\frac{1}{2}\frac{m_W^2}{v^2}
									\sum_j^3\mathcal{R}_{1j}\im[\kappa_f^j]
										\int_0^1dz\Bigg\{
										\frac{4-z}{z}-\frac{m_{H^\pm}^2-m_{H^0_j}^2}{m_W^2}
										\Bigg\}(1-z)
									C^{WH^\pm}_{WH^0_j}(z)
		,\end{align}
		\normalsize
and the quark-loop contributions to CEDM are
	%%%   BZ CEDM
		\small
		\begin{align}
				d_q^C(q')	&=\frac{m_q}{(16\pi^2)^2}4g_3^3
											\frac{m_{q'}^2}{v^2}
									\sum_j^3\int_0^1dz\Bigg\{
												\im[\kappa_q^j]\re[\kappa_{q'}^j]
												\rbra{\frac{1}{z}-2(1-z)}
												+\re[\kappa_q^j]\im[\kappa_{q'}^j]
												\frac{1}{z}
											\Bigg\}
									C^{gH^0_j}_{q'q'}(z)
		,\end{align}
		\normalsize
where $V=\gamma$, $Z$ and $N_C$ is the color factor.
The coupling constants are given by
$g_{\gamma ff}^v=eQ_f$,
$g_{Z ff}^v=g_z(I_f/2-Q_fs_W^2)$,
$g_{H^+ H^- \gamma}=-ie$,
$g_{H^+ H^- Z}=-ig_Z c_{2W}/2$,
$g_{H^\pm H^\mp H^0_j}=(\lam_3\mathcal{R}_{1j}+\re[\lam_7]\mathcal{R}_{2j}-\im[\lam_7]\mathcal{R}_{3j})v$,
$g_{WWA}=-ie$ and
$g_{WWZ}=-ig_Zc_W^2$,
where $Q_f$ are the electric charges of the fermion, $g_Z=\sqrt{g_1^2+g_2^2}$, $s_W=\sin\theta_W$, $c_W=\cos\theta_W$ and $c_{2W}=\cos2\theta_W$.
	%%%
		\begin{align}
				C^{GH}_{XY}(z)	&=C_0\rbra{0,0;m_G^2,m_H^2,\frac{(1-z)m^2_X+zm_Y^2}{z(1-z)}}
		,\end{align}
and where $C_0$ is the Passarino-Veltman function\cite{Passarino:1978jh},
	%%%
		\begin{align}
			C_0(0,0;m_1,m_2,m_3)=\frac{1}{m_1^2-m_2^2}
									\cbra{
										\frac{m_1^2}{m_1^2-m_3^2}\log\rbra{\frac{m_3^2}{m_1^2}}
										-\frac{m_2^2}{m_2^2-m_3^2}\log\rbra{\frac{m_3^2}{m_2^2}}
										}
		.\end{align}
We confirmed the consistency of our results with Refs.~\cite{Leigh:1990kf,BowserChao:1997bb,Jung:2013hka,Abe:2013qla,Cheung:2014oaa,Cheung:2020ugr}.

\section{Decay of the heavy Higgs bosons}\label{sc:decayrate}
We here list the partial decay width of the extra Higgs bosons given in Eqs.~\eqref{eq:decaymodes}, \eqref{eq:decaymodes3} and \eqref{eq:decaymodespm} in Sec.~\ref{sc:collider}.
We refer to Refs.~\cite{Bian:2017jpt,Djouadi:2005gi,Djouadi:2005gj,Christova:2002uja}.
	%%%   partial decay width
		\begin{align}
			\Gamma(H^0_i\to f\bar{f})	
				&=	\frac{N_C G_{F} m_{H^0_i} m_f^2}{4\sqrt{2}\pi}
					\left(1-\frac{4m_f^2}{m_{H^0_i}^2}\right)^{3/2}
					\left[
						(\re[\kappa_f^i])^2
						+(\im[\kappa_f^i])^2\left(1-\frac{4m_f^2}{m_{H^0_i}^2}\right)^{-1}
					\right]
			,\\
			\Gamma(H^0_i\to gg)
				&=	\frac{G_F \alpha_s^2 (m_{H^0_i}) m_{H^0_i}^3}{64 \sqrt2 \pi^3}
					\left[
						\left| \sum_q \re[\kappa_q^i] A_{1/2}^{H} (\tau^i_q) \right|^2
						+\left| \sum_q \im[\kappa_q^i] A_{1/2}^{A} (\tau^i_q) \right|^2
					\right]
			,\\
			\Gamma(H^0_i \to \phi V^*)
				&=	\frac{9 G_\mu^2 m_V^4}{8 \pi^3}
					\delta_V' m_{H^0_i} g_{H^0_i \varphi V}^2
					\,G\rbra{\frac{m_\varphi^2}{m_{H^0_i}^2},\frac{m_V^2}{m_{H^0_i}^2}}
			,\\
			\Gamma(H^+\to ff')
				&=	\frac{N_C}{16\pi m_{H\pm}^3}
					\sbra{
						\rbra{m_{H^\pm}^2-m_f^2-m_{f'}^2}^2-4m_f^2m_{f'}^2
					}^{1/2}
				\nn\\&\quad\times
					\sbra{
						\rbra{m_{H^\pm}^2-m_f^2-m_{f'}^2}\rbra{|y^+_f|^2+|y^+_{f'}|^2}
						-4m_fm_{f'}\re[y^+_f{}^*y^+_{f'}]
					}
		,\end{align}
where
$\tau^i_X=m_{H^0_i}^2/4m_X^2$,
$\delta_Z'=\frac{7}{12}-\frac{10}{9}\sin^2\theta_W+\frac{40}{9}\sin^4\theta_W$,
$\delta'_W=1$
and $\phi=H^0_{j\neq i}(H^\pm)$ for $V^*=Z^*(W^*)$.
The couplings
$y^+_f$ and $y^+_{f'}$
are given by
$\mathcal{L}\supset (\bar{f}_R y^+_f f'_L+\bar{f}_L y^+_{f'} f'_R) H^+ +h.c.$.
The functions $A(\tau)$ are given by
	%%%
		\eq{
				A_{1/2}^H(\tau)&=	2[\tau+(\tau-1)f(\tau)]\tau^{-2}
			,\\	A_{1/2}^A(\tau)&=	2\tau^{-1}f(\tau)
		,}
and where
	%%%
		\eq{
			f(\tau)=
				\left\{
				\begin{array}{ll}
					\arcsin^2\sqrt{\tau},
								&\quad\textrm{for}\quad	\tau\leq1
					,\\
					-\frac{1}{4}\sbra{\log\frac{1+\sqrt{1-\tau^{-1}}}{1-\sqrt{1-\tau^{-1}}}-i\pi}^2,
								&\quad\textrm{for}\quad	\tau>1
				.\end{array}
				\right.
		}
In terms of $\lambda(x,y)=-1+2x+2y-(x-y)^2$ with $x=m_X^2/m_{H^0_i}^2$,
the function $G(x,y)$ is given by
	%%%
		\begin{align}
			G(x,y)&=
				\frac{1}{4}
				\left\{
					2(-1+y-x)\sqrt{\lambda(x,y)}
					\left[ \frac{\pi}{2}+\arctan\left(\frac{y (1-y+x)-\lambda(x,y)}{(1-x) \sqrt{\lambda(x,y)}} \right) \right]
				\right.
				\nn \\
				&\quad\quad\quad
				\left.
					+(\lambda(x,y)-2x)\log x +\frac{1}{3}(1-x)\left[ 5(1+x) -4y -\frac{2}{y} \lambda(x,y) \right]
				\right\}.
		\end{align}

\section{Beta functions}\label{sc:betafunc}
We here list the beta functions of the dimensionless coupling constants of general THDM up to one-loop level given in Sec.~\ref{sc:RGE}.
The couplings are expressed in any basis of the fermions and the scalar doublet fields given in Eqs.~\eqref{eq:potential1} and \eqref{eq:yukawa1} in Sec.~\ref{sc:model}.
The beta functions of the gauge coupling constants are given as
	%%%
		\begin{align}
			16\pi^2\beta_{g_3}&=	\rbra{-11+\frac{4}{3}n_g}g_3^3
		,\\	16\pi^2\beta_{g_2}&=	\rbra{-\frac{22}{3}+\frac{4}{3}n_g+\frac{1}{6}n_d}g_2^3
		,\\	16\pi^2\beta_{g_1}&=	\rbra{\frac{20}{9}n_g+\frac{1}{6}n_d}g_1^3
		,\end{align}
where $n_g$ is the number of the generation of the fermions and $n_d$ is the number of scalar doublets.
The beta functions of the Yukawa-coupling matrices are given as
	%%%
		\begin{align}
			16\pi^2\beta_{y_{u,k}}=&	-2\sum_{l}^2(y_{u,l}y_{d,k}y_{d,l}^\dagger)+\frac{1}{2}\sum_l^2y_{u,k}[(y_{u,l}^\dagger y_{u,l})+(y_{d,l}y_{d,l}^\dagger)]+\sum_l^2(y_{u,l}y_{u,l}^\dagger)y_{u,k}
							\nn	\\&	+\sum_l^2\Tr[yy]_{lk}y_{u,l}
									-(8g_3^2+\frac{9}{4}g_2^2+\frac{17}{12}g_1^2)y_{u,k}
		,\\	16\pi^2\beta_{y_{d,k}}=&	-2\sum_{l}^2(y_{u,l}^\dagger y_{u,k}y_{d,l})+\frac{1}{2}\sum_l^2[(y_{u,l}^\dagger y_{u,l})+(y_{d,l}y_{d,l}^\dagger)]y_{d,k}+\sum_l^2y_{d,k}(y_{d,l}^\dagger y_{d,l})
							\nn	\\&	+\sum_l^2\Tr[yy]_{lk}y_{d,l}
									-(8g_3^2+\frac{9}{4}g_2^2+\frac{5}{12}g_1^2)y_{d,k}
		,\\	16\pi^2\beta_{y_{e,k}}=&	\frac{1}{2}\sum_l^2(y_{e,l}y_{e,l}^\dagger)y_{e,k}+\sum_l^2y_{e,k}(y_{e,l}^\dagger y_{e,l})
									+\sum_l^2\Tr[yy]_{lk}y_{e,l}
									-(\frac{9}{4}g_2^2+\frac{15}{4}g_1^2)y_{e,k}
		,\end{align}
where
	%%%
		\begin{align}
			\Tr[yy]_{lk}=\Tr[N_C\rbra{y_{u,l}^\dagger y_{u,k}+y_{d,l}^\dagger y_{d,k}}+y_{e,l}^\dagger y_{e,k}]
		.\end{align}
The beta functions of the scalar self-couplings are given as
	%%%
	\newcommand{\TrYYYY}[2]{{\Tr}\big[y_{#1,#2}^{}y_{#1,#2}^\dagger y_{#1,#2}^{}y_{#1,#2}^\dagger\big]}
	\newcommand{\TrYYYYf}[5]{{\Tr}\big[y_{#1,#2}^{}y_{#1,#3}^\dagger y_{#1,#4}^{}y_{#1,#5}^\dagger\big]}
	\newcommand{\TrYYYYg}[4]{{\Tr}\big[y_{u,#1}^{}y_{d,#2}^{}y_{d,#3}^\dagger y_{u,#4}^\dagger\big]}
		\small
		\begin{align}
			16\pi^2\beta_{\lam_1}=&~	2\cbra{6\lam_1^2+2\lam_3^2+2\lam_3\lam_4+\lam_4^2+|\lam_5|^2+12|\lam_6|^2}
							\nn	\\&	-4\cbra{N_C\rbra{\TrYYYY{u}{1}+\TrYYYY{d}{1}}+\TrYYYY{e}{1}}
							\nn	\\&	+4\Tr[yy]_{11}\lam_1+2\Tr[yy]_{21}(\lam_6+\lam_6^*)
									+\frac{3}{4}(3g_2^4+g_1^4+2g_2^2g_1^2)
									-3(3g_2^2+g_1^2)\lam_1
		,\\	16\pi^2\beta_{\lam_2}=&~	2\cbra{6\lam_2^2+2\lam_3^2+2\lam_3\lam_4+\lam_4^2+|\lam_5|^2+12|\lam_7|^2}
							\nn	\\&	-4\cbra{N_C\rbra{\TrYYYY{u}{2}+\TrYYYY{d}{2}}+\TrYYYY{e}{2}}
							\nn	\\&	+4\Tr[yy]_{22}\lam_2+2\Tr[yy]_{12}(\lam_7+\lam_7^*)
									+\frac{3}{4}(3g_2^4+g_1^4+2g_2^2g_1^2)
									-3(3g_2^2+g_1^2)\lam_2
		,\\	16\pi^2\beta_{\lam_3}=&~	2\cbra{(\lam_1+\lam_2)(3\lam_3+\lam_4)+2\lam_3^2+\lam_4^2+|\lam_5|^2+2|\lam_6|^2+2|\lam_7|^2+4(\lam_6\lam_7^*+\lam_6^*\lam_7)}
							\nn	\\&	-4\left\{N_C\left(\TrYYYYg{1}{2}{2}{1}+\TrYYYYg{2}{1}{1}{2}-\TrYYYYg{1}{2}{1}{2}-\TrYYYYg{2}{1}{2}{1}\right)\right.
									\nn\\&\quad\quad~+N_C\left.\left(\TrYYYYf{u}{2}{1}{1}{2}+\TrYYYYf{d}{2}{2}{1}{1}\right)+\TrYYYYf{e}{2}{2}{1}{1}\right\}
							\nn	\\&	+2(\Tr[yy]_{11}+\Tr[yy]_{22})\lam_3+\Tr[yy]_{12}(\lam_6+\lam_6^*)+\Tr[yy]_{21}(\lam_7+\lam_7^*)
							\nn	\\&	+\frac{3}{4}(3g_2^4+g_1^4-2g_2^2g_1^2)
									-3(3g_2^2+g_1^2)\lam_3
		,\\	16\pi^2\beta_{\lam_4}=&~	2\cbra{(\lam_1+\lam_2+4\lam_3)\lam_4+2\lam_4^2+4|\lam_5|^2+5|\lam_6|^2+5|\lam_7|^2+(\lam_6\lam_7^*+\lam_6^*\lam_7)}
							\nn	\\&	-4\left\{N_C\left(\TrYYYYg{2}{1}{2}{1}+\TrYYYYg{1}{2}{1}{2}-\TrYYYYg{2}{1}{1}{2}-\TrYYYYg{1}{2}{2}{1}\right)\right.
									\nn\\&\quad\quad~+N_C\left.\left(\TrYYYYf{u}{2}{1}{1}{2}+\TrYYYYf{d}{2}{2}{1}{1}\right)+\TrYYYYf{e}{2}{2}{1}{1}\right\}
							\nn	\\&	+2(\Tr[yy]_{11}+\Tr[yy]_{22})\lam_4+\Tr[yy]_{12}(\lam_6+\lam_6^*)+\Tr[yy]_{21}(\lam_7+\lam_7^*)
							\nn	\\&	+3g_2^2g_1^2
									-3(3g_2^2+g_1^2)\lam_4
		,\\	16\pi^2\beta_{\lam_5}=&~	2\cbra{(\lam_1+\lam_2+4\lam_3+6\lam_4)\lam_5+5\lam_6^2+5\lam_7^2+2\lam_6\lam_7}
							\nn	\\&	-4\cbra{N_C\rbra{\TrYYYYf{u}{2}{1}{2}{1}+\TrYYYYf{d}{2}{1}{2}{1}}+\TrYYYYf{e}{2}{1}{2}{1}}
							\nn	\\&	+2(\Tr[yy]_{11}+\Tr[yy]_{22})\lam_5+2\Tr[yy]_{12}\lam_6+2\Tr[yy]_{21}\lam_7
									-3(3g_2^2+g_1^2)\lam_5
		,\\	16\pi^2\beta_{\lam_6}=&~	2\cbra{(6\lam_1+3\lam_3+4\lam_4)\lam_6+(3\lam_3+2\lam_4)\lam_7+\lam_5(5\lam_6^*+\lam_7^*)}
							\nn	\\&	-4\cbra{N_C\rbra{\TrYYYYf{u}{1}{1}{2}{1}+\TrYYYYf{d}{1}{1}{2}{1}}+\TrYYYYf{e}{1}{1}{2}{1}}
							\nn	\\&	+(3\Tr[yy]_{11}+\Tr[yy]_{22})\lam_6+\Tr[yy]_{12}\lam_1+\Tr[yy]_{21}(\lam_3+\lam_4+\lam_5)
									-3(3g_2^2+g_1^2)\lam_6
		,\\	16\pi^2\beta_{\lam_7}=&~	2\cbra{(6\lam_2+3\lam_3+4\lam_4)\lam_7+(3\lam_3+2\lam_4)\lam_6+\lam_5(\lam_6^*+5\lam_7^*)}
							\nn	\\&	-4\cbra{N_C\rbra{\TrYYYYf{u}{2}{2}{2}{1}+\TrYYYYf{d}{2}{2}{2}{1}}+\TrYYYYf{e}{2}{2}{2}{1}}
							\nn	\\&	+(\Tr[yy]_{11}+3\Tr[yy]_{22})\lam_7+\Tr[yy]_{21}\lam_2+\Tr[yy]_{12}(\lam_3+\lam_4+\lam_5)
									-3(3g_2^2+g_1^2)\lam_7
		.\end{align}
		\normalsize
We confirmed the consistency of our results with Ref.~\cite{Branco:2011iw,Gori:2017qwg,Bijnens:2011gd}.

\end{document}